\documentclass[reprint,twocolumn,superscriptaddress,preprintnumbers,showpacs,nofootinbib,notitlepage,secnumarabic,amssymb, nobibnotes, aps, prd]{revtex4-2}

\usepackage{graphicx}
\usepackage{latexsym,amsmath,amssymb,lmodern,float,url}
\usepackage{dcolumn}
\usepackage{bm}
 \usepackage[normalem]{ulem} 
\usepackage{braket}
\usepackage{bbm}
\usepackage{amsthm}
\usepackage{natbib}
\usepackage{comment}

\usepackage[colorlinks=true,backref=false, linktocpage=true,
citecolor=blue,urlcolor=blue,linkcolor=blue,pdfpagemode=UseOutlines]{hyperref}
\usepackage[]{cleveref}

\hypersetup{%
  bookmarksnumbered=true,
  pdftitle = {},
  pdfsubject = {},
  pdfauthor = {},
  pdfkeywords = {}
}

\let\Re\undefined

\newcommand{\tr}{\mathrm{Tr}}
\DeclareMathOperator{\Tr}{Tr}
\DeclareMathOperator{\Re}{Re}

\DeclareMathOperator{\re}{Re}

\newcommand{\ds}{\mathbb{V}}

\newcommand{\bi}{\mathbb{BI}}

\newcommand{\sh}[1]{{\color{red} #1}}

\begin{document}
\preprint{FERMILAB-PUB-21-364-PPD}
\title{Primitive Quantum Gates for Dihedral Gauge Theories}
\author{M. Sohaib Alam}
\email{malam@usra.edu}
\affiliation{Rigetti Computing, Berkeley, CA, 94701, USA}
\affiliation{Quantum Artificial Intelligence Laboratory (QuAIL), NASA Ames Research Center, Moffett Field, CA, 94035, USA}
\affiliation{USRA Research Institute for Advanced Computer Science (RIACS), Mountain View, CA, 94043, USA}
\author{Stuart Hadfield}
\email{stuart.hadfield@nasa.gov}
\affiliation{Quantum Artificial Intelligence Laboratory (QuAIL), NASA Ames Research Center, Moffett Field, CA, 94035, USA}
\affiliation{USRA Research Institute for Advanced Computer Science (RIACS), Mountain View, CA, 94043, USA}
\author{Henry Lamm}
\email{hlamm@fnal.gov}
\affiliation{Fermi National Accelerator Laboratory, Batavia, IL, 60510, USA}
\author{Andy C. Y. Li}
\email{cli@fnal.gov}
\affiliation{Fermi National Accelerator Laboratory, Batavia,  IL, 60510, USA}

\collaboration{SQMS collaboration}

\date{\today}

\begin{abstract}
We describe the simulation of dihedral gauge theories on digital quantum computers. 
The nonabelian discrete gauge group $D_N$ -- the dihedral group -- serves as an approximation to $U(1)\times\mathbb{Z}_2$ lattice gauge theory. In order to carry out such a lattice simulation, we detail the construction of efficient quantum circuits to realize basic primitives including the nonabelian Fourier transform over 
$D_N$, the trace operation, 
and the group multiplication and inversion operations. For each case 
the required quantum resources 
scale linearly or as low-degree polynomials in $n=\log N$. 
We experimentally benchmark our gates on the Rigetti Aspen-9 quantum processor for the case of $D_4$. The estimated fidelity of all $D_4$ gates was found to exceed $80\%$.
\end{abstract}

\maketitle

\section{\label{sec:introduction}Introduction}
A promising area for quantum advantage is simulating the dynamics of nonperturbative quantum field theories~\cite{Feynman:1981tf,Lloyd1073,Jordan:2011ne,Jordan:2017lea,klco2021standard}. In order to propagate for a time $t$, one requires the unitary operator $U(t)=e^{-iHt}$ which in general may be challenging to efficiently implement on a quantum computer. Different quantum algorithms exist for approximating $U(t)$, in particular, 
Trotter-Suzuki product formulas~\cite{Jordan:2011ne,Jordan:2011ci,Jordan:2017lea,Garcia-Alvarez:2014uda,Jordan:2014tma,Moosavian:2017tkv,Bender:2018rdp,haah2018quantum,Du:2020glq,PhysRevX.11.011020,PhysRevLett.123.070503},
quantum walks~\cite{berry2009black}, 
Taylor series approximations~\cite{PhysRevLett.114.090502},  
and quantum signal processing~\cite{Low2019hamiltonian,PhysRevLett.118.010501}, as well as more recent variational approaches~\cite{cirstoiu2020variational,gibbs2021longtime,yao2020adaptive}.  While each of these algorithms differs in how to approximate $U(t)$, fundamentally these methods all require implementing operations derived from the Hamiltonian~$H$ as quantum circuits~\cite{Lamm:2019bik}. Thus, a small set of 
primitive 
operations should be required for all of them. In the case of gauge theories, the Kogut-Susskind Hamiltonian $H_{KS}$~\cite{PhysRevD.11.395} is the most common Hamiltonian discussed in the literature for quantum  simulations. Using $H_{KS}$, initial comparisons between a few quantum algorithms was performed for the Schwinger model~\cite{Shaw:2020udc}. 

For efficient digital simulations, the local lattice degrees of freedom must be truncated. For fermionic degrees of freedom, this is relatively easy~\cite{Jordan:1928wi,Bravyi2002FermionicQC,Chen:2018nog}. Further proposals discuss how to map lattice fermions (e.g. Wilson and staggered) to these encodings~\cite{Muschik:2016tws} or use gauge symmetry to eliminate them~\cite{Zohar:2018cwb,Zohar:2019ygc}. The question of gauge boson digitization is murkier, with many proposals~\cite{Zohar:2012ay,Zohar:2012xf,Zohar:2013zla,Zohar:2014qma,Zohar:2015hwa,Zohar:2016iic,Klco:2019evd,Ciavarella:2021nmj,Bender:2018rdp,Liu:2020eoa,Hackett:2018cel,Alexandru:2019nsa,Yamamoto:2020eqi,Haase:2020kaj,Armon:2021uqr,PhysRevD.99.114507,Bazavov:2015kka,Zhang:2018ufj,Unmuth-Yockey:2018ugm,Unmuth-Yockey:2018xak,Kreshchuk:2020dla,Kreshchuk:2020aiq,Raychowdhury:2018osk,Raychowdhury:2019iki,Davoudi:2020yln,Wiese:2014rla,Luo:2019vmi,Brower:2020huh,Mathis:2020fuo,Singh:2019jog,Singh:2019uwd,Buser:2020uzs,Bhattacharya:2020gpm,Barata:2020jtq,Kreshchuk:2020kcz,Ji:2020kjk,Gustafson:2021qbt} that make complicated tradeoffs. Digitizing reduces symmetries -- either explicitly or through finite-truncations~\cite{Zohar:2013zla}. Furthermore, the utility of a given scheme depends upon spacetime dimensionality~\cite{Zohar:2021nyc}. Care must be taken, as the regulated theory may not have the original theory as its continuum limit~\cite{Hasenfratz:2001iz,Caracciolo:2001jd,Hasenfratz:2000hd,PhysRevE.57.111,PhysRevE.94.022134,car_article}.

One promising digitization method is the approximation of gauge theories by discrete subgroups~\cite{Bender:2018rdp,Hackett:2018cel,Alexandru:2019nsa,Yamamoto:2020eqi,Ji:2020kjk,Haase:2020kaj,Carena:2021ltu,Armon:2021uqr}. Replacing the continuous group by a discrete subgroup was explored in the early days of Euclidean lattice field theory as a resource reduction procedure, with most studies focusing on the theories in $(3+1)$ dimensions with the Wilson action. The viability of the $\mathbb{Z}_N$ subgroups replacing $U(1)$ were studied in~\cite{Creutz:1979zg,Creutz:1982dn}. Further studies of the crystal-like discrete subgroups of $SU(N)$ were performed~\cite{Bhanot:1981xp,Petcher:1980cq,Bhanot:1981pj,Hackett:2018cel,Alexandru:2019nsa,Ji:2020kjk}, including with fermions~\cite{Weingarten:1980hx,Weingarten:1981jy}. Alongside this work, theoretical studies revealed that such discrete subgroup approximations correspond to effective field theories of continuous groups where a mass is given to the gauge fields through the Higgs mechanism~\cite{Kogut:1980qb,romers2007discrete,Fradkin:1978dv,Harlow:2018tng,Horn:1979fy}.  The result of this mass is that the discrete subgroup fails to well approximate the continuous group below a certain lattice spacing $a_f$ (or equivalently beyond a certain coupling $\beta_f$).

In lattice calculations, one performs calculations at fixed lattice spacing $a=a(\beta)$ which shrinks as $\beta\rightarrow\infty$ for asymptotically free theories. To control extrapolation errors in taking $a\rightarrow0$, one simulates in the \emph{scaling regime} of $a\ll m_{IR}^{-1}$ where $m_{IR}$ is the infrared mass scale of the physics of interest. We will consider the start of the scaling regime as occurring at $a_s$. Thus,
the approximation error 
from using discrete subgroups should be small provided that $a_s\gtrsim a_f$ or equivalently that $\beta_s\lesssim\beta_f$. For the Wilson action, $\beta_f$ are known. In the case of $U(1)$ in $(3+1)$-$d$ with $\beta_s= 1$, $\mathbb{Z}_{n>5}$ satisfies $\beta_f>\beta_s$. 
For nonabelian groups, only a finite set of crystallike subgroups exist. $SU(2)$ has three: the binary tetrahedral $\mathbb{BT}$, the binary octahedral $\mathbb{BO}$, and the binary icosahedral $\bi$. 
While $\mathbb{BT}$ has $\beta_f=2.24(8)$ in $3+1d$, $\mathbb{BO}$ and $\bi$ have $\beta_f=3.26(8)$ and $\beta_f=5.82(8)$ respectively~\cite{Alexandru:2019nsa}, above $\beta_s=2.2$. Hence, $\mathbb{BO}$ and $\bi$ appear useful for $SU(2)$.

For $SU(3)$ (the theory underlying QCD) with $\beta_s\approx6$ in $(3+1)$-$d$, all five crystal-like subgroups have $\beta_f < \beta_s$ , with the largest, the 1080-element Valentiner group\footnote{Sometimes called $S(1080)$~\cite{Bhanot:1981pj,Flyvbjerg:1984dj,Flyvbjerg:1984ji,Alexandru:2019nsa} or $\Sigma_{3\times360}$~\cite{hagedorn2014mixing}
.}, $\ds$ having $\beta_f=3.935(5)$~\cite{Alexandru:2019nsa}. Thus the discrete approximation is inadequate when using the Wilson action. By extending $\ds$ to include the midpoints between elements of $\ds$, one can increase $\beta_f\approx7$~\cite{Lisboa:1982jj}. However this require more qubits and sacrifices gauge symmetry.  This gauge violation is dangerous on noisy quantum computers~\cite{Stryker:2018efp,Halimeh:2019svu,1797835}.  An alternative approach to decrease $a_f$ introduces additional terms into the lattice action~\cite{Edgar:1981dr,Bhanot:1981pj,Creutz:1982dn,Fukugita:1982kk,Horn:1982ef,Flyvbjerg:1984dj,Flyvbjerg:1984ji,Ayala:1989it,Alexandru:2019nsa}, although only in~\cite{Bhanot:1981pj,Alexandru:2019nsa} were Monte Carlo calculations undertaken for $SU(3)$.  In \cite{Alexandru:2019nsa} it was shown that such modified actions of $\ds$ could reach into the scaling regime, finding calculations could be undertaken at $a>0.08$ fm without the effects of $a_f$ being seen. This suggest that $\ds$ can reproduce $SU(3)$ in the scaling region with a modified action, such that practical quantum computations of $SU(3)$ could be performed.

Nonabelian gauge theories have a number of novel features not seen in abelian ones, and thus studies of abelian theories like $U(1)$ or 
$\mathbb{Z}_N$ may be unrepresentative of the full complexity of lattice gauge theories. Unfortunately, even the smallest crystal-like subgroup of a nonabelian theory, $\mathbb{BT}$ requires 6 qubits per register and is thus beyond current hardware.  To reduce this cost to be more inline with near-term devices, in this work we study a class of discrete groups that are not crystal-like subgroups of a single continuous group.  The binary dihedral groups $D_N$ have $2N$ elements and are each an extension of 
$\mathbb{Z}_N$ by an additional $\mathbb{Z}_2$ subgroup giving 
$D_N\simeq \mathbb{Z}_N\rtimes\mathbb{Z}_2$. 
In the limit of $N\rightarrow\infty$ this becomes $U(1)\times\mathbb{Z}_2$.  
$D_3$ and $D_4$ have previously been investigated for simulation on quantum computers~\cite{Bender:2018rdp,Lamm:2019bik}.  Having 6 and 8 elements respectively, they both require 3 qubits per register.  Unfortunately in both $(2+1)$-$d$ and $(3+1)$-$d$, these two groups have $a_s<a_f$ with the standard Wilson action and thus either a modified action or larger group is required to minimize the discrete group approximation error.  For larger $N$, the necessary primitive gates are unknown, and within this work we will derive a set of such gates for $D_{2^n}$ gauge theories which naturally map onto qubit devices.  

\begin{figure}[ht]
\centering
    \includegraphics[width=\linewidth]{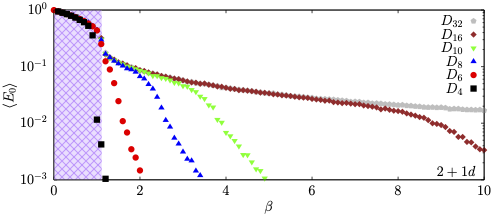}
        \includegraphics[width=\linewidth]{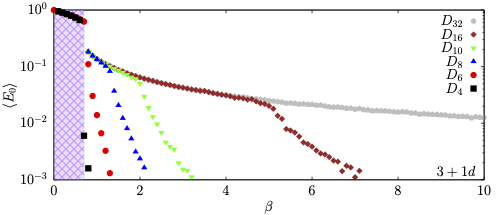}
    \caption{\label{fig:freezing}Euclidean calculations of lattice energy density $\langle E_0\rangle$ as measured by the expectation value of the plaquette as a function of Wilson coupling beta on $4^d$ lattices for different $D_N$ groups (top) $d=2+1$ dimensions and (bottom) $d=3+1$ dimensions. The shaded region corresponds to couplings outside the scaling regime for the $U(1)\times\mathbbm{Z}_2$ theory.} 
\end{figure}

Since we are interested in finding gauge theories that could be simulated on near-term quantum devices, it behooves us to study not just $(3+1)$-$d$ but also consider $(2+1)$-$d$ theories. Using classical lattice simulations, we have determined that in both spacetimes, while $\beta_f>\beta_s$ for $D_8$, it is only slightly larger, so either $D_{9}$ (which we did not simulate) or $D_{10}$ would be desirable to have simulations with sufficiently small $a$ (See Fig.~\ref{fig:freezing}).  The dependence of $a_f$ on $\beta_f$ within the scaling regime is exponential, so a slightly larger group can have dramatically smaller errors. Since $D_{2^n}$ theories can more efficiently be implemented in qubits, we believe that the 5-qubit $D_{16}$ should be the ultimate target for quantum hardware of the near-future, with $D_4$ and $D_8$ as important stepping stones to it.  After this, $\mathbb{BT}$ would be a natural next step.  

In this paper, we construct quantum circuits 
implementing the four primitive gates (inversion, multiplication, trace, and Fourier) required to simulate the $D_{2^n}$ theories. A Trotterized time evolution circuit can be built using these gates. Although Trotterized evolution on quantum processors is unfeasible at present due to the limited two-qubit gate fidelities, we benchmark the primitive gates for $D_4$ individually on the Rigetti Aspen-9 QPU to evaluate if simulations are practical on near-term quantum processors.
We perform process tomography to measure the process fidelities of the trace and Fourier gates. Since process tomography is experimentally costly for a six-qubit gate, the fidelity of the multiplication gate is estimated by the fraction of the correct bit-strings produced for all possible pairs of input bit-strings. The process fidelity of the inversion gate is reported based on the benchmark result of the CCPHASE gate \cite{hill2021realization}.

This paper is organized as follows. In Sec.~\ref{sec:lattice} the Euclidean action lattice formalism is briefly reviewed and its connection to the Hamiltonian formulation is elucidated. Sec.~\ref{sec:coherences} presents an overview of the four primitive gates required for implementing the group operations necessary for lattice gauge theories on quantum computers. This is followed by
quantum circuit constructions for these gates for $D_{2^n}$ 
gauge theories: 
the inversion gate in Sec.~\ref{sec:inverse}, the multiplication gate in Sec.~\ref{sec:multiplication}, the trace gate in Sec.~\ref{sec:trace}, and the Fourier transform gate in Sec.~\ref{sec:Fourier}. Benchmarking results for our $D_4$ gates on the Rigetti Aspen-9 QPU are found in Sec.~\ref{sec:results}.  We conclude and discuss future work in Sec.~\ref{sec:Conclusions}.

\section{\label{sec:lattice}Lattice Field Theory}
In order to understand the origin of the various primative gates requires from simulating lattice gauge theory, it is useful to review the connection between the Kogut-Susskind Hamiltonian~\cite{PhysRevD.11.395} and the Euclidean Wilson action.  We summarize the derivation of~\cite{Creutz:1984mg} that begins with the anisotropic Wilson action in Euclidean time $\tau=it$ defined on a spacetime lattice:
\begin{equation}\label{eq:Wilson}
    S_E=-\beta_t\sum_{t}\re\Tr U_t-\beta_s\sum_{s}\re\Tr U_s
\end{equation}
where $i= t,s$ refers to temporal and spatial plaquettes $U_i$ formed from gauge links. We introduce anisotropy between the lattice spacings by using different bare couplings on the spatial and temporal plaquettes:
\begin{align}
\label{eq:betas}
    \beta_t(a,a_0)=\frac{a}{g^2_t(a,a_0) a_0},\qquad    \beta_s(a,a_0)=\frac{a_0}{g^2_s(a,a_0) a}.
\end{align}

The renormalized anisotropic parameter $\xi \equiv a/a_0$ is used to denote the physical change in the lattice spacings caused by tuning the bare parameters $\beta_t,\beta_s$. To approach the Hamiltonian limit ($a_0\rightarrow0)$, it becomes natural to introduce two new couplings, $g_H^2=g_s g_t$, and the speed of light, $c=g_s g_t^{-1}$.

The connection to the Hamiltonian is via the transfer matrix, $T(a,a_0)$ which takes a state at time $\tau$, $|\tau\rangle$, to $|\tau+1\rangle$.  $T$ is related to $S_E$ by the partition function $Z$:
\begin{align}
    Z=\int D Ue^{-S_E}=\tr T(a_0)^N
\end{align}
where $N$ counts time slices. It follows that the matrix elements of $T(a_0)$ are~\cite{Creutz:1984mg}
\begin{align}\label{eq:transfer}
    \langle \tau+&1|T(a_0)|\tau\rangle\nonumber\\=&e^{\frac{\beta_s}{2}\sum_{s}\re \tr  U_s}e^{\beta_t\sum_{\{\tau,\tau+1\}} \re \tr U_t}e^{\frac{\beta_s}{2}\sum_{s}\re \tr  U_s}\notag\\
    \equiv&T_{V}^{1/2}T_KT_{V}^{1/2},
\end{align}
where we have used the second-order Trotterization. Since $T(a_0)\equiv e^{-a_0 H(a,a_0)}$, we desire $T(a_0)$ in terms of variables on one time slice. While this is trivial for $U_s$, $U_t$ couples links at two times $U_{ij}(\tau),U_{ij}(\tau+1)$. To proceed, we gauge-fix in the temporal gauge, $U_{0i}=\mathbbm{1}$, yielding
\begin{equation}
    S_{K}=-\beta_t\sum_{\{\tau,\tau+1\}}\re\Tr U_{ij}(\tau)U_{ij}^\dag(\tau+1).
\end{equation}
The next step is to express $T(a_0)$ in terms of operators.  The link operator is simply $\hat{U}_{ij}|\tau\rangle={U}_{ij}|\tau\rangle$. For $T_K$, we need an operator that evolves a link via $R_{ij}(g)|\tau\rangle =|\tau'\rangle,$ where $U_{ij}\rightarrow g U_{ij}$.
This operator has the property of $R_{ij}(g)R_{ij}(h)=R_{ij}(gh)$ and can define a conjugate momentum to $\hat{U}_{ij}$ by performing a rotation on $U_{ij}(\textbf{x},\tau+1)$. With this, we write 
\begin{equation}
T_K=\prod_{\{ij\}}T_{K,ij}=\prod_{\{ij\}}\left[\int Dg R_{ij}(g) e^{\beta_t\re\tr g}\right],
\end{equation}
where the product is over all spatial links $U_{ij}(\tau)$.  Any group element equals $g=e^{i\omega\cdot\lambda}$ where $\lambda_i$ are the adjoint generators, and $R_{ij}(g)=e^{i\omega\cdot l_{ij}}$ in terms of the generators $l_{ij}$ of that representation.  Defining $\prod_\alpha(D\omega^\alpha)J(\omega)$ as the invariant group measure with a Jacobian $J$, $T_K(a_0)$ is
\begin{align}
\label{eq:tkijaprox}
    T_{K,ij}=\int \prod_\alpha(D\omega^\alpha)J(\omega) e^{i\omega\cdot l_{ij}} e^{\beta_t\tr \cos(\omega\cdot \lambda)}.
\end{align}
Summing over all character functions of the group, this integral is can be performed analytically, requiring the Fourier transform over the group. This was used in~\cite{Lamm:2019bik} and seems to be a viable procedure when the group is finite. On the other hand, when dealing with continuous groups, the summation contains 
infinite character functions and is thus computationally impractical.

To remedy this obstacle in continuous groups, one expands $T_K$ to $O(\omega^2)$ leaving Gaussian integrals.
Neglecting the overall normalization, $T(a_0)$ is
\begin{equation}\label{eq:Ta0}
    T(a_0)=e^{\frac{\beta_s}{2}\sum_s\re\tr \hat{U}_s}e^{-\beta_t^{-1}\sum_{\{ij\}}l_{ij}^2}e^{\frac{\beta_s}{2}\sum_s\re\tr \hat{U}_s}.
\end{equation}
Regardless of whether one approximates $T_K$ or not, the integral corresponds to the Fourier transform of the gauge group.  This transform which diagonalizes the kinetic energy is why we need a primative quantum Fourier transform gate for the given group.

From Eq.~(\ref{eq:Ta0}), we use the definition of $T(a_0)$ to obtain a lattice Hamiltonian. However, since $l_{ij}$ and $\hat{U}_{ij}$ do not commute, rearranging this expression into a single exponential requires application of the Baker-Campbell-Hausdorff (BCH) formula, yielding:
\begin{widetext}
\begin{align}
\label{eq:latham}
    H(a,a_0)=&\frac{1}{c(a,a_0)a}\bigg(g_H^2(a,a_0)\sum_{\{ij\}}l_{ij}^2-g^{-2}_H(a,a_0)\sum_s\re\tr \hat{U}_s\notag\\
    &-\frac{1}{24}\frac{1}{c^2(a,a_0)\xi^2}\sum_{\{ij\},s}\left(g_H^{2}(a,a_0)[2l_{ij}^2,[l_{ij}^2,\re\tr \hat{U}_s]]-g_H^{-2}(a,a_0)[\re\tr\hat{U}_s,[l_{ij}^2,\re\tr \hat{U}_s]]\right)+\hdots\bigg).
\end{align}
\end{widetext}

Taking the $a_0\rightarrow0$ limit of $T(a_0)$:
\begin{equation}
    \mathcal{T}(\tau)\equiv \lim_{a_0\rightarrow 0, N\rightarrow\infty}T(a_0)^N,
\end{equation}
the BCH terms vanish and one obtains $H_{KS}\equiv-\frac{1}{\tau}\log(\mathcal{T}(\tau))$~\cite{PhysRevD.11.395}:
\begin{equation}
\label{eq:ksham}
    H_{KS}=\frac{1}{c(a)a}\left(g_H^2(a)\sum_{\{ij\}}l_{ij}^2-\frac{1}{g_H^2(a)}\sum_{s}\re\Tr U_s\right).
\end{equation}
This is a common starting point for the evolution of lattice gauge theories on quantum computers.  From this, we see that in order to simulate these gauge theories, there are a number of basic, group-dependent gates that can be used to simulate the two terms of Eq.~(\ref{eq:ksham}). Along with the quantum Fourier transform for the kinetic term, for the potential term we need to be able to introduce a phase $\propto \Re\Tr U_s$.  This is the origin of the need for the trace gate, the multiplication gate (needed to compute the plaquette $U_s$ from the four links forming it), and the inversion gate (needed to uncompute the plaquette).

\section{\label{sec:coherences}Overview of Basic Gates}
In ref.\ \cite{Lamm:2019bik}, the basic set of gates requires for a general gauge group $G$ were given.  This construction begins with the defining for $G$ a qubit $G$-register by identifying each group element with a computational basis state $\ket{g}, g\in G$. 
For pure-gauge Hamiltonians, a set of useful
primitive gates defined on the $G$-register are: inversion, multiplication, trace, and the quantum Fourier transform. 

The inversion gate acts on a single $G$-register mapping each group element to its inverse. 
This is defined in the fiducial basis by
\begin{equation}
\mathfrak U_{-1} \left|g\right> = \left|g^{-1}\right>\text.
\end{equation}


The group (matrix) multiplication gate acts on two $G$-registers and is defined by

\begin{equation}
\mathfrak U_\times \left|g\right> \left|h\right> = \left|g\right> \left|gh\right>\text.
\end{equation}
Here we have defined $\mathfrak U_\times$ as implementing in-place left multiplication,
in the sense that the content of the second register was multiplied on the left. 
Left multiplication suffices for a minimal set as right multiplication can be implemented using two applications each of 
$\mathfrak U_{-1}$ and $\mathfrak U_\times$~\cite{Lamm:2019bik}, although fidelity of simulations may be improved by directly implement right multiplication as well.

The trace of a plaquette appears in $H_{KS}$, and so to perform this operation we combine the matrix multiplication gate with a single-register trace gate:
\begin{equation}
\mathfrak U_{\Tr}(\theta) \left|g\right> = e^{i \theta \Re\Tr g} \left|g\right>.
\label{eqn:trace-gate}
\end{equation}


In our construction, the final gate required on the $G$-register is the Fourier transform gate $\mathfrak U_F$. This gate acts on a $G$-register to rotate into 
the Fourier basis:
\begin{equation}
\mathfrak U_F \sum_{g \in G} f(g)\left|g \right>
=
\sum_{\rho \in \hat G} \hat f(\rho)_{ij} \left|\rho,i,j\right>.
\end{equation}
The second sum is taken over $\rho$, the representations of $G$, and $\hat f$ denotes the Fourier transform of $f$. This gate diagonalizes what will be the `kinetic' part of the Trotterized time-evolution operator. After application of the gate, the register is no longer a $G$-register but a $\hat G$-register.

In the subsequent section we consider quantum circuit implementations of these gates for the dihedral group $D_N=\{g=s^mr^k|s^2=r^N=e\}$ generated by a reflection $s$ and rotation $r$; 
we review the important properties of $D_N$ in Appendix~\ref{appendix:algebraic-DN}. 
Following \cite{Lamm:2019bik}, the $2N$ group elements $s^mr^{k}$, $m\in\{0,1\}$, $k \in \{0, \dots, N-1\}$, are encoded using standard binary in the qubit computational basis states $\vert m \rangle \vert k \rangle$, where the register $\vert k\rangle$ uses 
$\left\lceil \log_2{N} \right\rceil$ qubits. We may variously refer to the $\vert m \rangle$ as the $s$-qubit or the reflection qubit, and the $\vert k \rangle$ as the $r$-register or the rotation register. In this paper, we focus exclusively on the case $N=2^n$, so that in all we need $n + 1$ qubits  
to encode all the elements of $D_N$.

\section{\label{sec:inverse}Inversion Gate}

Here we describe how to construct a circuit realizing the inversion gate $\mathfrak U_{-1}  \ket{g}=\ket{g^{-1}}$ for $D_N$. 
First, consider the case of a general discrete gauge group $G$. 
As observed in \cite{Lamm:2019bik}, 
if we have access to both the multiplication gate~$\mathfrak U_{\times}$ and its reversed (adjoint) circuit $\mathfrak U_{\times}^\dagger$, 
then we can implement
~$\mathfrak U_{-1}$ 
using an ancillary $G$-register initialized to the group identity element~$\ket{e}$. We 
discuss construction of the multiplication gate 
in Sect. \ref{sec:multiplication}. 
We can then implement $\mathfrak U_{-1}$ using the 
sequence of operations
\begin{equation} \ket{g}\ket{e}\xrightarrow{\mathfrak U_{\times}^\dagger} \ket{g}\ket{g^{-1}} \xrightarrow{SWAP} \ket{g^{-1}}\ket{g}\xrightarrow{\mathfrak U_{\times}}\ket{g^{-1}}\ket{e},
\end{equation}
at which point the ancillary register has been returned to $\ket{e}$ and 
can be reused or discarded. 
We note that the SWAP may be performed virtually by simply switching (relabelling) the top and bottom registers in the circuit for~$\mathfrak U_{\times}$. 
Hence the cost of this implementation of ~$\mathfrak U_{-1}$ is at most twice that of~$\mathfrak U_{\times}$. 
Note that the property that the ancilla register is initialized and returned to a fixed state 
can be used to further simplify the circuits 
for~$\mathfrak U_{\times}^\dagger$ so that fewer gates are required than for the general case. 
In any case, the use of ancilliary $G$-register means this implementation requires at least 
$\log |G|$ additional qubits, with $\log |G|=n+1$ for $D_N$. 

Alternatively, 
one may use the properties of $D_N$ (see Appendix \ref{appendix:algebraic-DN}) to derive more specific constructions requiring 
fewer ancilla qubits and lower circuit depth. 
For this we use 
that the inverse of an element $s^{m}r^{k}$ is given by 
\begin{equation}
    \left( s^{m} r^{k} \right)^{-1} =  s^{m} r^{(N-k)(1-m) + mk}
\label{eqn:inverse}
\end{equation}
As a result, given the qubit encoding $s^{m}r^{k} \rightarrow \vert m \rangle \vert k \rangle$ described above, 
the effect of $\mathfrak U_{-1}$ is to change the state of the register $\vert 0 \rangle \vert k \rangle \rightarrow \vert 0 \rangle \vert N-k \rangle$, and leave $\vert 1 \rangle \vert k \rangle$ unmodified. Therefore, controlled on the state of the left-most qubit, we need to compute the 2's complement of the register $\vert k \rangle$. The 2's complement of an $n$-bit binary number is defined as its complement with respect to $2^n$, so that the sum of the number and its complement equals $N=2^n$ \sh{$(\equiv 0)$.}
It can be obtained by first taking its 1's complement, i.e., flipping all the 0s to 1s and 1s to 0s, and then adding 1 to the resulting integer. 

\begin{figure}[h]
\centering
    \includegraphics[width=7cm]{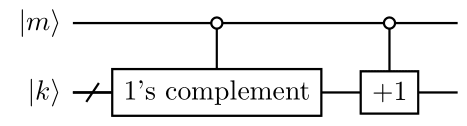}
    \caption{Schematic quantum circuit implementation of the inversion gate~$\mathfrak U_{-1}$. The controlled 1's complement operation can be implemented with $n$ CNOT gates. The controlled increment gate can be implemented with $O(n)$ Tofolli gates using a constant number of additional ancilla qubits~\cite{gidney15,haner2016factoring}.} 
    \label{fig:invert_schem}
\end{figure}

\begin{figure}[h]
\includegraphics[width=7cm]{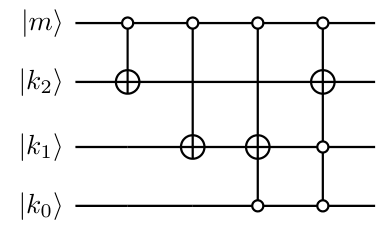}
    \caption{Example: A simplified implementation of the inversion gate $\mathfrak U_{-1}$ for $D_8$. The first two gates correspond to the 1's complement operation, and the last two to incrementation; an inner pair of CNOTs has canceled out.} 
    \label{fig:invert_d8}
\end{figure}

Hence, controlled on $m=0$, we apply a Pauli $X$ gate (i.e., a CNOT) to each qubit in $\ket{r}$, followed by the increment operation. Treating the register $\ket{r}$ as an integer mod $N$, the increment operation can be implemented using simplified versions of standard quantum circuits for addition; 
various constructions with different tradeoffs in terms of size, depth, and number of ancilla qubits can be found in the literature, 
see in particular \cite[Table 1]{haner2016factoring}. 
In terms of circuit depth, a straightforward modification of the constructions of~\cite{gidney15,haner2016factoring} yields quantum circuits for the controlled-increment operation using 
$O(n)$ Tofolli gates and as few as $1$ additional ancilla qubits. 
%
 %
A schematic circuit for the inversion gate is shown in Fig.~\ref{fig:invert_schem}, and an example circuit is shown in Fig.~\ref{fig:invert_d8}. 

\section{\label{sec:multiplication}Multiplication Gate}

Here we describe how to build a circuit realizing the multiplication gate $\mathfrak U_{\times} \vert g \rangle \vert h \rangle = \vert g \rangle \vert gh \rangle$ for the dihedral group. 
We employ the following group multiplication rule for elements of $D_N$
\begin{eqnarray}
s^{m_1} r^{k_1} \cdot s^{m_2} r^{k_2} &=& s^{m_1 + m_2} r^{Nm_2 + (-1)^{m_2}k_1 + k_2}, 
\label{eqn-product}
\end{eqnarray}
which implies that it suffices
to construct a circuit that performs either addition or subtraction depending on whether $m_2 = 0$ or $m_2 = 1$. Therefore, the task of realizing the $D_N$ multiplication gate reduces to performing 
conditional binary arithmetic on qubits.

In the case where $m_2=0$, we must add $k_1$ and $k_2$, whereas in the case where $m_2=1$, we must add the two's complement of $k_1$ and $k_2$. The construction of the circuit to compute the two's complement is given in the section on the inversion gate \ref{sec:inverse}, except that now we control on the value of the leading qubit being 1 instead of 0. Having conditionally prepared the two's complement of $k_1$, we must then perform binary addition to complete the computation of Eq. \eqref{eqn-product}.


A variety of proposed quantum 
algorithms for addition and multiplication exist in the literature~\cite{hadfield2016scientific,haner2018optimizing} with different resource tradeoffs, see, e.g., \cite[Table 1]{Takahashi10}. 
One approach is to use the 
classical full-adder, which takes inputs $A$ and $B$, the two bits to be added, and $C_{in}$, the carry-in bit from the previous bit addition, and outputs the sum $S$ and the carry-out $C_{out}$. In Reed-Muller form, these are given by
\begin{eqnarray}
S &=& A \oplus B \oplus C_{in} \nonumber \\
C_{out} &=& AB \oplus AC_{in} \oplus BC_{in}.
\end{eqnarray}

If we choose to over-write one of the registers with the sum, say the register containing the $A$ bits in the convention above, then we we can compute $S$ at every step using 2 CNOTs, one controlled on the value of $B$ and the other on the value of $C_{in}$, with the target being $A$. Similarly, we can compute $C_{out}$ and write out its value to an ancillary qubit at every step using 3 CCNOT gates. Therefore, for a $D_{2^n}$ gauge theory, using this scheme we would require 2 CNOTs to compute the sum and 3 CCNOTs to compute the carry outs for each of $n-2$ bits, in addition to $n-1$ ancillary qubits to hold the value of the carries. We would only need 1 CNOT to compute the sum of the least significant bit, and 1 CCNOT to compute the carry-out for this bit. We also do not need to compute the carry-out of the most significant bit. Assuming 1 CCNOT $\sim$ 6 CNOTs, in all this adds a cost of $20n - 31$ CNOTs in addition to the circuit to compute the two's complement in order to implement the multiplication gate.

\begin{figure}[h]
    \includegraphics[width=8cm]{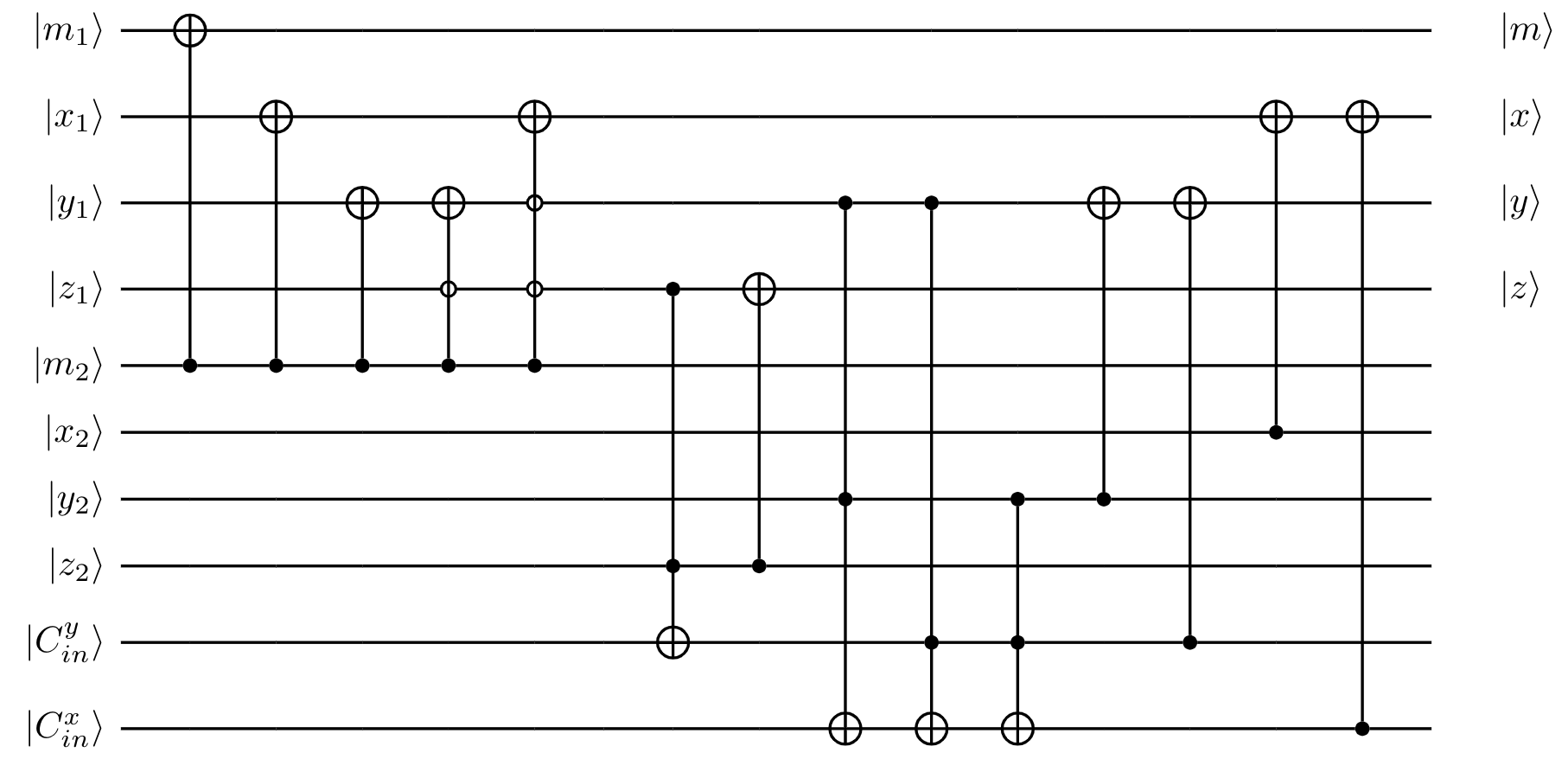}
    \caption{A multiplication gate circuit for $D_8$ theory}
    \label{fig:d8mult}
\end{figure}

An example implementation for the $D_8$ multiplication gate is shown in Fig. \ref{fig:d8mult}.

\section{\label{sec:trace}Trace Gate}

Here we describe how to construct quantum circuits realizing the trace gate $\mathfrak U_{\Tr}(\theta) \ket{g}=e^{i\theta \Re \Tr g} \ket{g}$ for 
$D_N$. Observe that, unlike the other basic gates we consider, this family of gates is parameterized by a real number $\theta$. We describe both a straightforward implementation that scales with $N$, and is in principle exact, as well as a more complicated implementation that scales polynomially with $n:=\log_2 N$, 
but generally comes with 
some degree of approximation error; we refer to 
these as \textit{direct} and \textit{ancilla-assisted} implementations, respectively. 
Though its implementation cost has worse asymptotic scaling, the direct construction may 
be especially useful when $N$ is relatively small, such as the important cases 
for near-term experiments
described in Section~\ref{sec:introduction}. 

Here $\Tr g$ corresponds to the matrix trace in the fundamental representation. We let $H_{\Tr}$ denote the diagonal Hamiltonian defined as $H_{\Tr}\ket{g}= \Re(\Tr(g))\ket{g}$ such that 
 $\mathfrak U_{\Tr}(\theta)\ket{g}=e^{i\theta \Re(\Tr(g))}=e^{i\theta H_{\Tr}}$.  

For $D_N$, in the fundamental (two-dimensional) representation for each group element $g=s^m r^k$ 
we have
\begin{equation} \label{eq:fundRep}
\rho(g) = \begin{pmatrix}
0 & 1 \\
1 & 0  
\end{pmatrix}^m
 \begin{pmatrix}
\omega & 0 \\
0 & \overline{\omega}
\end{pmatrix}^k, \;\;\;\;\textrm{ where }\;\; \omega = e^{2\pi i/N}.
\end{equation}
Clearly $\rho(g)$ is always traceless when $m=1$. 
When $m=0$ we have $\Tr(g):=\Tr(\rho(g))=\omega^k + \omega^{-k}=2\cos(2\pi k /N),$ 
and so for each $N$ the trace values 
uniformly 
sample from one period of $2\cos(x)$. 
Hence for $D_N$ we observe $\Re\Tr(g)=\Tr(g)$. Therefore we have 
\begin{equation}  \label{eq:HamTrace}
     H_{\Tr} = \ket{0}\bra{0}\otimes \sum_{k =0}^{N-1} 2\cos(2\pi k/N) \ket{k}\bra{k}.
\end{equation}
\paragraph*{Direct implementation:} We first consider implementation of $\mathfrak U_{\Tr}(\theta)$ by directly simulating evolution under the Hamiltonian $H_{\Tr}$ for time~$\theta$. 
Diagonal Hamiltonians on $n$ qubits can be uniquely written as $H=\sum_{\alpha}a_\alpha Z_\alpha$, where $Z_\alpha=Z_{\alpha_1}\dots Z_{\alpha_j}$ denotes a tensor product of Pauli $Z$ operators indexed by subsets of qubits $\alpha\subset [n]$, with coefficients given by $a_\alpha =\tfrac1{2^{n}}\Tr [Z_\alpha H]\in \mathbb{R}$~\cite{hadfield2018representation}. 
We may apply this decomposition to $H_{\Tr}$, or, to take advantage of its tensor structure, to only the second factor on the right-hand side of  Eq.~(\ref{eq:HamTrace}), as desired;
in general, however, the number of non-zero terms in such a decomposition is proportional to $N$. Nevertheless, for moderate $N$ this decomposition yields an straightforward implementation of $\mathfrak U_{\Tr}(\theta)$. For the latter case, i.e., writing $H_{\Tr}=\ket{0}\bra{0}\otimes \sum_\alpha a_\alpha Z_\alpha$, we have  
\begin{eqnarray}
\mathfrak U_{\Tr}(\theta)=\prod_{\alpha=0}^N e^{i\theta a_\alpha \ket{0}\bra{0}\otimes Z_\alpha}
=\prod_{\alpha=0}^N \Lambda_{m=0} (e^{i\theta a_\alpha Z_\alpha}),
\end{eqnarray}
where $\Lambda_{m=0} (e^{i\theta a_\alpha Z_\alpha})$ denotes the controlled unitary implementing $e^{i\theta a_\alpha Z_\alpha}$ conditioned on the first qubit being zero, and we used the fact that diagonal terms mutually commute. Each controlled rotation can be implemented with $O(n)$ basic gates consisting of CNOTs and single-qubit gates~\cite{barenco1995elementary,hadfield2018representation}; however the number of rotations may be proportional to $N=2^n$. An advantage of this approach is that if qubit rotations can be implemented exactly then so can $\mathfrak U_{\Tr}(\theta)$. If we tolerate approximation error in $\mathfrak U_{\Tr}(\theta)$ the gate costs can be further reduced~\cite{welch2014efficient}.

For example, consider $D_4$, for which Eq.~(\ref{eq:fundRep}) gives 
$\rho(g) = X^m(iZ)^k$, 
which has trace $2$ for $g=e$ and $-2$ for $g=r^2$, else $0$. Hence we have the Hamiltonian
$H_{\Tr}=2\ket{000}\bra{000}-2\ket{010}\bra{010}$
which we may write as
$H_{\Tr}=\frac12(Z_{k_1}+Z_mZ_{k_1}+Z_{k_1}Z_{k_0}+Z_{m}Z_{k_1}Z_{k_0})$,
or with control as 
$H_{\Tr}=\ket{0}\bra{0}\otimes (Z_{k_1}+Z_{k_1}Z_{k_0})=2\ket{0}\bra{0}\otimes Z_{k_1}\otimes \ket{0}\bra{0}.$
So for $D_4$ we see that we can implement $U_{\Tr}$ exactly with a double-controlled $Z$ rotation, or a controlled $Z$ and controlled $ZZ$ rotation, or a combination of $Z$, two $ZZ$, and a $ZZZ$ rotations. 


\paragraph*{Ancilla-enabled implementation:} 
On the other hand, when $N$ becomes large it is desirable to have a quantum circuit for the trace gate with resource costs that scales polynomially with $n$ as opposed to $N=2^n$.  
This can be accomplished if we accept tradeoffs such as the use of ancilla qubit registers and some degree of approximation error. Here the basic idea is that we use the ancilla registers as scratchpad space for quantum arithmetic circuits 
that coherently compute the trace value for each group element, upon which we apply controlled rotation gates to achieve the desired phase kickback. Clearly, for real numbers any finite size ancilla register will lead to some degree of approximation error, in general, in the computed values and resulting phases. This error may be systematically reduced by employing larger ancilla registers and circuits that utilize higher precision numbers; we leave a detailed analysis of these time, space, and precision tradeoffs for future work.  

Let's first consider restricting the required trigonometric quantities to the first 
first quadrant $0\leq 2\pi \ell /N < \pi/2$ which will simplify construction of the resulting quantum circuits. In particular, many approximations for computing 
numerical functions 
come with guaranteed precision only over such a bounded interval, and moreover some trigonometric algorithms proceed by computing values of $\cos$ and $\sin$ simultaneously. Observe that for each group element $\ket{g}=\ket{mk_{n-1}\dots k_1 k_0}$ we have $\Tr(g)=2(1-m)\cos(2\phi k/N)$, and so 
the periodicity of the cosine function implies
that the bit $k_{n-1}$ controls the sign of the coefficient and the bit $k_{n-2}$ controls the 'phase', i.e., explicitly 
$\Tr(0k_{n-1}0k_{n-3}\dots k_1 k_0) = 2(-1)^{k_{n-1}}\cos(2\pi k' /N)$ and 
$\Tr(0k_{n-1}1k_{n-3}\dots k_1 k_0) = -2(-1)^{k_{n-1}}\sin(2\pi k /N)$, 
where $k'$ 
is the integer given by the bits $k_{n-3}\dots k_1 k_0$. 
(Note that a similar treatment of the first 3 bits may be employed in the direct case above for the $D_4$ example.)


Assume for the moment we can implement the desired quantum arithmetic modules for computing fixed-precision trigonometric functions to $b$ bits of accuracy after the decimal point~\cite{hadfield2016scientific}. Then $\mathfrak U_{\Tr}$ can be implemented as follows:
\begin{enumerate}
    \item (Compute classical functions.)
    Append a sufficiently large ancilla register $\ket{00\dots 0}$ and reversibly compute (in superposition) the transformation for each basis state $\ket{g}=\ket{mk_{n-1}k_{n-2}k'}$
    \begin{align}
    &\ket{mk_{n-1}k_{n-2}k'}\ket{0\dots 0}\notag\\&\rightarrow \ket{mk_{n-1}k_{n-2}k'}\ket{\widetilde{\sin(2\pi k'/N)}}\ket{\widetilde{\cos(2\pi k'/N)}}\ket{sp},\notag\end{align}
    where $\widetilde x$ denotes a $b$-bit 
    binary approximation of a quantity $0\leq x<1$. 
    The remaining scratchpad register,  $\ket{sp}$, denotes intermediate classical values 
    which will be used to facilitate uncomputation. 
   We discuss how this may be implemented below.
   
    \item (Phase kickback.) 
    Given a $b$-bit quantity $0\leq x<1$  we can implement $\ket{x}\rightarrow e^{i \theta x}\ket{x}$ (up to an 
    irrelevant global phase) using a controlled $R_Z(\theta 2^{-j})$ gate applied to each $j$th bit of $\ket{x}$, $j=1,\dots,b$, such that the number of such gates is $b$. 
    The single-qubit $Z$ rotation gate is defined as $R_Z(\phi)=e^{-i\phi Z/2}$. 
    Similarly, we can implement $\ket{x}\rightarrow e^{i 2\theta x}\ket{x}$ applying instead controlled $R_Z(\theta 2^{1-j})$ gates. 
    Hence we apply two high-level unitaries that kickback a phase of $\theta Tr(g)$ to each basis state, as schematically depicted in Fig.~\ref{fig:traceIntermediate}.
\begin{figure}[h]
\includegraphics[width=\linewidth]{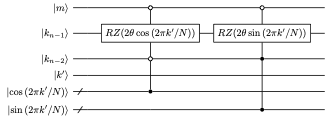}
    \caption{Schematic for phase kickback of $\Tr(g)$ using registers containing the required $\cos$ and $\sin$ values. Here the two schematic gates represent one multicontrolled-Z rotation for each bit in the $\sin$/$\cos$ registers, respectively, i.e., the gate $R_Z(\theta 2^{1-j})$ is applied for each $j$th bit, $j=1,\dots,b$, encoding the $2^{-j}$ bit.} 
    \label{fig:traceIntermediate}
\end{figure}    
    
        \item (Uncomputation.) As the operation of the second step is diagonal, we can restore the ancilla qubits to $\ket{00\dots 0}$ then discard for reuse by applying the reverse 
        of the circuit in Step 1. Hence each input basis state is taken to the desired state
        $$\ket{g}=\ket{mk} \rightarrow e^{i\theta 2(1-m)\cos(2 \pi k /N)} \ket{mk}= e^{i\theta \Tr(g)} \ket{g}$$
\end{enumerate}

The cost of Steps $1$ and $3$ depend on the arithmetic subroutine used and the number $b$ of bits of accuracy in the $\sin$ and $\cos$ registers. This cost dominates that of Step $2$ which depends 
linearly on
~$b$. 
Here we've assumed the ancilla qubits are restored to $\ket{00\dots 0}$ 
for reuse; 
we note that the allocation and uncomputation of ancilla qubit resources may often be significantly optimized within the context of an overall algorithm~\cite{hadfield2016scientific}.

\paragraph*{Computing the trigonometric functions:} Methods for computing the $\cos$ and $\sin$ functions using quantum arithmetic circuits are discussed in  \cite{cao2013quantum,hadfield2016scientific,haner2018optimizing,wang2020quantum}. Different approaches come with different tradeoffs in terms of the number of qubits, number and types of basic gates, and required numerical accuracy for a given application. 

The approach of \cite[Sec. 5 and App. 2]{cao2013quantum} requires only addition and multiplication operations, and simultaneously computes both $\sin$ and $\cos$ using repeated squaring via the approximation
\begin{align} e^{i \theta a}&=\cos(a) + i\sin(a)=(e^{i \theta a/R})^R\notag\\& \,\simeq\, (1-i\theta a/R - (\theta a/R)^2/2)^R\end{align}
such that the quantity $1-i\theta a/R - (\theta a/R)^2/2$ is computed by storing separately its real and imaginary parts, for a suitable $R=2^r\gg 1$ selected with respect to the accuracy parameter~$b$.  Repeatedly squaring this quantity  (requiring only $r$ operations) then yields the desired $\sin$ and $
\cos$ approximations. Roughly, the error in these approximations goes as $1/R$ for sufficiently many bits of accuracy in the $\sin$ and $\cos$ registers as well as the intermediate quantities (cf. \cite[Prop. 1 and 2]{cao2013quantum}).    

Alternatively, 
 the approach of \cite[App. D]{haner2018optimizing} uses piecewise polynomial approximations implemented via controlled Horner polynomial evaluation (such that each degree $d$ polynomial approximation requires $d+1$ additions and multiplications), while the approach of \cite{wang2020quantum} employs nontrivial quantum submodules for approximately computing square roots. In these approaches care must be taken to ensure the desired accuracy is achieved. 

Hence the ancilla-assisted approaches yield quantum circuits with resource costs scaling as low-degree polynomials in $n=\log N$ and the accuracy bits $b$. 
The specific cost in terms of gates and ancillas depends on these parameters and the particular quantum arithmetic circuits employed as subroutines. 
As stated the direct approach is much simpler for moderate $N$;  we show an explicit quantum circuit for 
this implementation of $\mathfrak U_{\Tr}(\theta)$ for $D_4$ and its compilation to hardware gates in Fig.~\ref{fig:circuit_trace_Fourier_d4} below. 



\section{\label{sec:Fourier}Fourier Gate}

The standard $n$-qubit quantum Fourier transform~\cite{nielsen_chuang_2010}, a critical component of Shor's prime factoring algorithm, corresponds to the abelian group $\mathbb{Z}_{2^n}$. 
Quantum circuits implementing Fourier transforms over a variety of nonabelian groups have been considered in~
\cite{hoyer1997efficient,beals1997quantum,puschel1999fast,moore2006generic}, though there remains important groups for which efficient QFT circuits are not known~\cite{childs2010quantum}.\footnote{We note that an efficient quantum circuit for the QFT of a group $G$ does not entail an efficient quantum algorithm for the corresponding Hidden Subgroup Problem (HSP) for $G$, an important class of problems that includes both the prime factoring and graph isomorphism problems~\cite{childs2010quantum}. Subexponential time quantum algorithms for the HSP on $D_N$ are given in \cite{regev2004subexponential,kuperberg2011another} 
using the standard QFT rather than the $D_N$ one considered here. 
}

Here we consider the explicit construction of quantum circuits for the QFT on $D_N$. Our construction employs the standard QFT as a subroutine. 
We note that the more general construction of \cite{hoyer1997efficient} for efficient circuits for QFTs over metacyclic groups also includes $D_N$.

The Fourier transform of a representation of some finite group $G$ is defined as
\begin{eqnarray}
\hat{f}(\rho) = \sqrt{\frac{d_{\rho}}{N}} \sum_{g \in G} f(g) \rho(g),
\label{eqn:Fourier-group}
\end{eqnarray}
where $N = \vert G \vert$, $d_{\rho}$ is the dimensionality of the representation $\rho$, and $f$ is a function over $G$. The inverse transform is given by
\begin{eqnarray}
f(g) = \frac{1}{\sqrt{N}} \sum_{\rho \in \hat{G}} \sqrt{d_{\rho}} \tr{(\hat{f}(\rho) \rho(g^{-1}))},
\end{eqnarray}
where the dual $\hat{G}$ is the set of all irreducible representations (irrep) of $G$. Note, if there exists a subgroup $H \subset G$ and elements $\{g_i\}_{i=1}^{n}$ such that we can write $G = \cup_{i=1}^{n} g_i H$, i.e., a \textit{left transversal} of $H$ exists in $G$, then
\begin{eqnarray}
\sum_{g \in G} f(g) \rho(g) &=& \sum_{i=1}^{n} \sum_{h \in H} f(g_i h) \rho(g_i h)  \\
&=& \sum_{i=1}^{n} \rho(g_i) \sum_{h \in H} f_i (h) \rho(h) = \sum_{i=1}^{n} \rho(g_i) \hat{f}_i (\rho \vert_H)\notag
\end{eqnarray}
where we have defined $f_i(h) = f(g_i h)$, $\rho \vert_H$ denotes the restriction of the representation $\rho$ to the subgroup $H$, and $\hat{f}_i$ represents the Fourier transform of the function $f_i$. Using this, we can compute the Fourier transform $\hat{f}$ on the representation $\rho$ in a recursive manner for the series of subgroups $H_1, \dots, H_n$ that form a chain $G \supset H_1 \supset \dots \supset H_n = {id}$, using ``adapted bases" such that $\rho \vert_{H_{i}}$ can be written as a direct sum of irreps of $H_{i}$.\\

Let $\alpha: G \rightarrow \mathbb{C}$ and $\alpha_i (g)\equiv \alpha(g_{i}g)$, where $g_i,g \in G$. Then, we may similarly construct quantum Fourier transforms (QFTs) via the following series of operations
\begin{eqnarray}
\vert \psi \rangle &=& \sum_{g \in G} \alpha (g) \vert g \rangle = \sum_{i=1} \sum_{h \in H} \alpha(g_i h) \vert g_i \rangle \vert h \rangle \nonumber \\
&=& \sum_{i=1}^{n} \vert g_i \rangle \left( \sum_{h \in H} \alpha_{i}(h) \vert h \rangle \right) \nonumber \\
&\xrightarrow{\text{$F_H$}}& \sum_{i=1}^{n} \vert g_i \rangle \left( \sum_{\tilde{h} \in \hat{H}} \hat{\alpha}_{i}(\tilde{h}) \vert \tilde{h} \rangle \right) \nonumber \\
&\xrightarrow{\text{$U$}}& \sum_{\tilde{g} \in \hat{G}} \hat{\alpha}(\tilde{g}) \vert \tilde{g} \rangle = \vert \tilde{\psi} \rangle
\label{eqn:generic-qft}
\end{eqnarray}
where $F_H$ denotes the Fourier transform over the subgroup $H$, and $U$ denotes a change of basis from $T \otimes B_H$ to $B_G$, where $T$ denotes the coset representatives $\{g_i \}_{i=1}^{n}$ and $B_H$ ($B_G$) denotes the Fourier basis of the group $H$ (respectively $G$). In our encoding of $D_N$ elements $s^{m} r^{k} \rightarrow \vert g \rangle = \vert m \rangle \vert k \rangle$, $\vert k \rangle$ encode the basis elements of $\mathbb{Z}_N$, while $\vert \tilde{k} \rangle$ denote the Fourier basis of $\mathbb{Z}_N$. Then, we have $F_H : \vert m \rangle \vert k \rangle \rightarrow \vert m \rangle \vert \tilde{k} \rangle$. Likewise, denoting the Fourier basis of $D_N$ by $\vert \tilde{g} \rangle$, the final transformation is $U: \vert m \rangle \vert \tilde{k} \rangle \rightarrow \vert \tilde{g} \rangle$. Determining $U$ is often the more non-trivial part of any such QFT algorithm.\\

\begin{figure}
    \includegraphics[width=6cm]{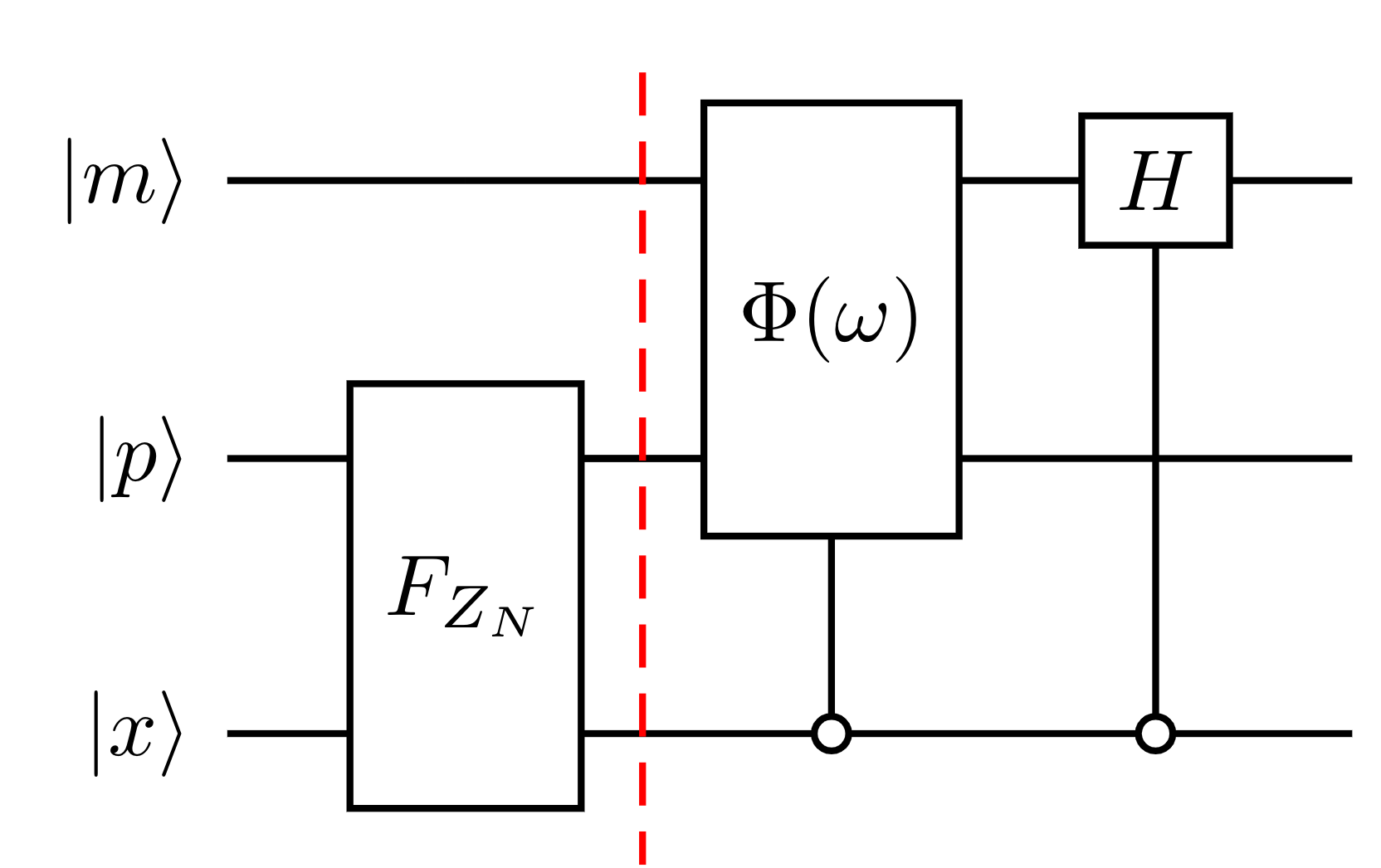}
    \caption{Circuit computing $\mathfrak U_F$ of the $D_N$ group. The first part of the circuit computes the Fourier transform $F_{\mathbb{Z}_N}$ over $\mathbb{Z}_{N} = \{r^{0}, \dots, r^{N-1}\}$, while the latter part performs a change of basis $\vert g_i \rangle \vert \tilde{h}\rangle \rightarrow \vert \tilde{g} \rangle$, implementing the unitary transform given in Eq. \eqref{eqn:qft-change-basis}.}
    \label{fig:dn-qft}
\end{figure}

\begin{figure}
    \centering
    \includegraphics[width=6cm]{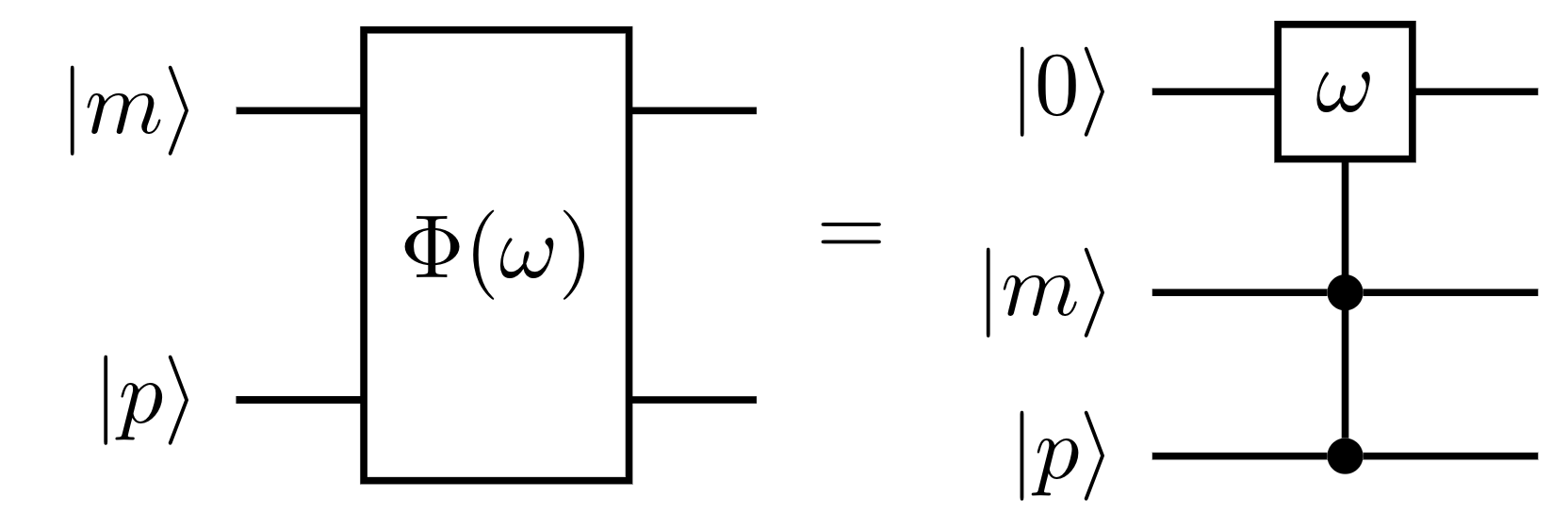}
    \caption{Circuit for implementing $\Phi(\omega)$. This is given by a simple application of a $\mathrm{CCPHASE}$ gate targeted on an ancillary qubit, obtaining a phase kick-back on the qubits $\vert k \rangle$ and $\vert i \rangle$, on which it is controlled.}
    \label{fig:Phi}
\end{figure}

For even $N$, the group $D_N$ has the following four 1-dimensional irreps using $m\in \{0,1\}$ and $k \in \{0\dots N-1\}$:
\begin{itemize}
    \item $\rho_A:$ $r^{k} \rightarrow 1$, $sr^{k} \rightarrow 1$
    \item $\rho_B:$ $r^{k} \rightarrow 1$, $sr^{k} \rightarrow -1$
    \item $\rho_C:$ $s^{m}r^{k} \rightarrow 1$ for even $k$; $s^{m}r^{k} \rightarrow -1$ for odd $k$
    \item $\rho_D:$ $r^{k} \rightarrow 1$, $sr^{k} \rightarrow -1$ for even $k$; and $r^{k} \rightarrow -1$, and $sr^{k} \rightarrow 1$ for odd $k$,
\end{itemize}
and $\frac{N-2}{2}$ 2-dimensional irrep:
\begin{eqnarray}
\phi^{(l)}(s^{m}r^{k}) = \begin{pmatrix}
0 & 1 \\
1 & 0
\end{pmatrix}^{m} \begin{pmatrix}
e^{i 2 \pi l/N} & 0 \\
0 & e^{-i 2 \pi l/N}
\end{pmatrix}^{k},
\label{eqn:DN-evenN-2d-irreps}
\end{eqnarray}
where $1 \leq l < \frac{N}{2}$. $D_N$ has a cyclic subgroup $Z_N = \{r^{0}, \dots, r^{N-1}\}$ for which the QFT is well known~\cite{nielsen_chuang_2010}, and for which the elements $\{e, s\}$ provide a left transversal in $D_N$. Our encoding of $D_N$ elements into qubits, $s^{m} r^{k} \rightarrow \vert m \rangle \vert k \rangle$, and the existence of the QFT over $Z_N$ provide all the steps in Eq. \eqref{eqn:generic-qft} to compute the QFT over $D_N$ except the last one involving a change of basis. This non-trivial step is provided by \cite{hoyer1997efficient}
\begin{widetext}
\begin{eqnarray}
U: \vert mN + p\frac{N}{2} + x \rangle &\rightarrow& \left\lbrace \begin{array}{cc}
\vert mN + p\frac{N}{2} + x \rangle, & 1 < x < \frac{N}{2} \\
\left( e^{i\pi \frac{N}{2}} \right)^{pm} \frac{1}{\sqrt{2}} \sum_{j=0}^{1} \left( e^{i\pi} \right)^{jm} \vert jN + p\frac{N}{2} + x \rangle, & x=0
\end{array} \right.
\label{eqn:qft-change-basis}
\end{eqnarray}
\end{widetext}
where $m,p \in \{0,1\}$ are the 2 most significant bits, while $x \in \{0,\dots,\frac{N}{2}-1\}$ specifies the state of the remaining part of the register. The complete circuit for $\mathfrak U_F$ for $D_N$ ($N=2^n$ for some $n$) is given in Fig. \ref{fig:dn-qft}. There, we use the operation $\Phi(\omega) \vert u \rangle \vert v \rangle = \omega^{uv} \vert u \rangle \vert v \rangle$, with $\omega = e^{i\pi\frac{N}{2}}$. In general, if $u$ takes on $n_1$ values and $v$ takes on $n_2$ values, then $\Phi$ can be compiled using $\Theta (\left\lceil log(n_1) \right\rceil \left\lceil log(n_2) \right\rceil)$ gates. In our case however, $m$ and $p$ only take on 2 values each, and we can therefore compile this operation using a single CCPHASE gate and an ancillary qubit, as shown in Fig.~\ref{fig:Phi}. With this formulation, we reduce the gate costs for $D_4$ from 5 entangling gates in~\cite{Lamm:2019bik} to 2.

Upon the execution of the Fourier gate, the four 1-dimensional irreps of $D_N$ (with $N = 2^n$) are encoded into the following basis states \begin{align}
 \rho_A \rightarrow \vert 00 \rangle \vert 0 \rangle^{\otimes n-1},&\hspace{0.5cm}
     \rho_B \rightarrow \vert 10 \rangle \vert 0 \rangle^{\otimes n-1}\notag\\
     \rho_C \rightarrow \vert 01 \rangle \vert 0 \rangle^{\otimes n-1},&\hspace{0.5cm}
 \rho_D \rightarrow \vert 11 \rangle \vert 0 \rangle^{\otimes n-1}\notag   
\end{align}
while the matrix entries $\rho_{ij}^{(l)}$, the $i$-th row and $j$-th column of the 2-dimensional irreps indexed by $l \in [1, \frac{N}{2})$ and given by Eq. \eqref{eqn:DN-evenN-2d-irreps} are encoded into the remaining computational basis states as $\rho_{ij}^{(l)} \rightarrow \vert ij \rangle \vert l \rangle$
If so desired, one could rearrange the representations to appear in a different order, e.g. the first four computational basis states $\vert \dots 00\rangle$, $\vert \dots 01\rangle$, $\vert \dots 10\rangle$, $\vert \dots 11\rangle$ encoding the four 1-dimensional irreps, the next four encoding the matrix entries of the $l=1$ 2-dimensional irrep and so on. In principle, the amplitudes of any two basis states $\vert s\rangle$ and $\vert s^{\prime} \rangle$ could be exchanged by using an ancillary qubit $\vert t \rangle$ and applying an $(n+1)$-qubit controlled operation $C^{n+1}(s) = \vert s \rangle \langle s\vert \otimes X + \sum_{s\neq s^{\prime}=0}^{2^{n+1}-1} \vert s^{\prime} \rangle \langle s^{\prime} \vert \otimes \mathbb{I}$ on a single target qubit $\vert t \rangle$, followed by at most $n+1$ CNOTs controlled on $\vert t \rangle$ to change $\vert s \rangle$ to $\vert s^{\prime} \rangle$.

However, for us, this is unnecessary since to apply the kinetic gate, we only ever need to apply the Fourier gate to transform to the momentum basis, and thereupon apply a diagonal operator, followed by the inverse of the Fourier gate to move back to position basis. In Appendix \ref{appendix:proof-Fourier-kinetic}, we prove that the Fourier gate diagonalizes the kinetic gate for $D_N$ theory (even $N$), satisfying our requirement.

\section{\label{sec:results}Experimental Results}
\begin{figure}
    \includegraphics[width=8cm]{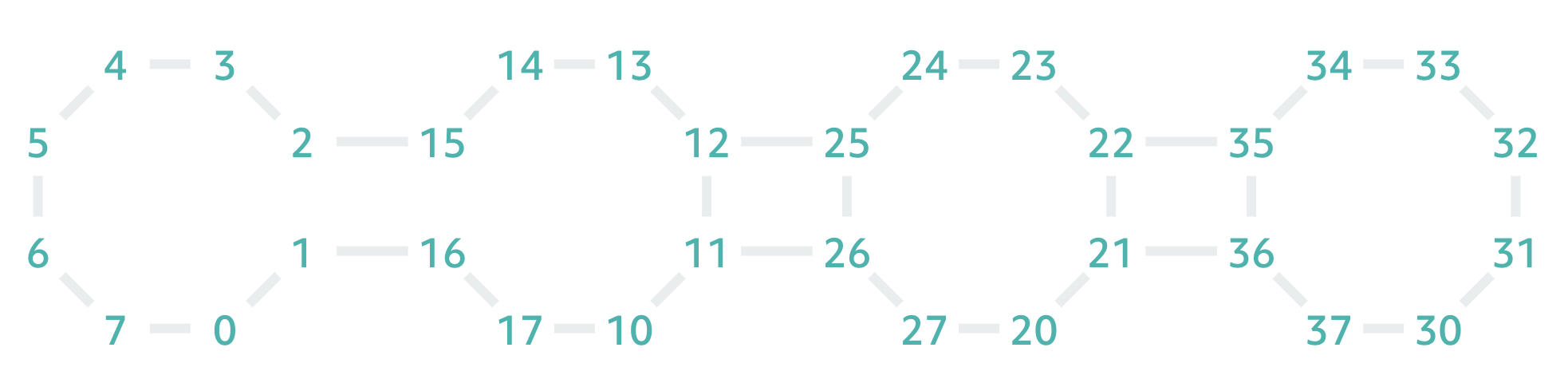}
    \caption{The Rigetti Aspen-9 QPU consists of 32 transmon qubits arranged in a square-octagon lattice. Single-qubit gates \{$\mathrm{RZ}(\theta)$, $\mathrm{RX}(k\pi/2)$\} are available for the qubits denoted by the integers. Native two-qubit gates (CPHASE, CZ, XY) can be applied between the connected qubits with the linkage denoted by grey lines.}
    \label{fig:aspen9}
\end{figure}

In this section, we discuss experimental results from running realizations of the circuits described above on the Rigetti Aspen-9 QPU, which features 32 transmon qubits with a square-octagon topology \cite{PhysRevA.101.012302,Abrams2020,Reagor2018} (see Fig.~\ref{fig:aspen9}). The Rigetti stack \cite{Karalekas_2020} allows us to use the Quil language \cite{smith2016practical} to program the Aspen-9 device, and its associated optimizing compiler Quilc \cite{smith2020opensource} to compile fundamental gates into its native gateset $\{\mathrm{RZ}(\theta), \mathrm{RX}(k\pi/2), \mathrm{CPHASE}, \mathrm{CZ}, \mathrm{XY}\}$. A recently realized native gate CCPHASE \cite{hill2021realization} is also accessible using the Quil language. We report the process fidelities of the Fourier, inversion and trace gates for $D_4$ theory. The multiplication gate for $D_4$ involves a 6-qubit circuit, performing process tomography on which is experimentally costly. Instead, we compute the fraction of correct bitstrings the gate produces for all possible pairs of input bitstrings, and report this as the \textit{accuracy} of this operation as a proxy to its fidelity. We find all the gates to have greater than $80\%$ fidelity or accuracy.

\subsection{\texorpdfstring{$D_{4}$}{D4} Multiplication Gate}\label{subsecn:d4_mult_gate}

Concretely, for $D_4$ theory, we can use the encoding $sr^j \rightarrow \vert s \rangle \vert j \rangle = \vert a\rangle \vert bc \rangle$ as in \cite{Lamm:2019bik} to specify an element of $D_4$ as $s^{a} r^{2b+c}$. We compute the product $\vert abc \rangle = \vert (a_1 b_1 c_1) \cdot (a_2 b_2 c_2) \rangle$, using the multiplication gate $\mathfrak U_{\times} \vert a_1 b_1 c_1 \rangle \vert a_2 b_2 c_2 \rangle = \vert a_1 b_1 c_1 \rangle \vert abc \rangle$. Whether we perform subtraction or addition, the right-most bit will simply be given by $c = c_1 \oplus c_2$. The relation is similar for the left-most bit ($a$).

For the second-right-most bit ($b$), we must first mod-2 sum both bits involved in the product, $b_1 \oplus b_2$. However, we must also account for the carry from (to) the mod-2 addition (subtraction) of the right-most bit. Depending on whether we perform addition or subtraction, the appropriate carry is either $c_1 c_2$ or $c_1 \bar{c}_2$ respectively. Thus, in all, we have the two following rules.\\

For $m_2=0$ (addition), we obtain
\begin{equation}
a = a_{1} \oplus a_{2} ,\hspace{0.3cm}
b = b_{1} \oplus b_{2} \oplus c_1 c_2 ,\hspace{0.3cm}
c = c_{1} \oplus c_{2}
\end{equation}

For $m_2=1$ (subtraction), we obtain the product
\begin{equation}
a = a_{1} \oplus a_{2} ,\hspace{0.3cm}
b = b_{1} \oplus b_{2} \oplus c_1 \bar{c}_2 ,\hspace{0.3cm}
c = c_{1} \oplus c_{2}
\end{equation}

\begin{figure}
    \includegraphics[width=7cm]{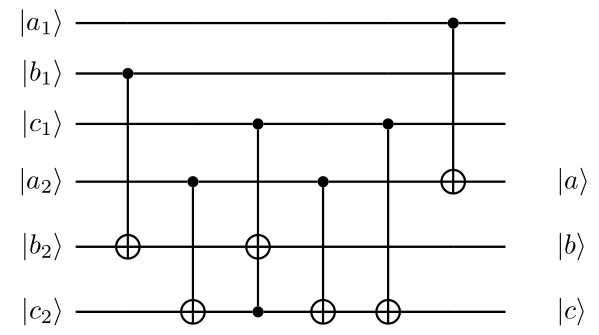}
    \caption{Multiplication gate circuit for $D_4$ theory}
    \label{fig:d4mult}    
\end{figure}

In circuit form, this is provided in Fig. (\ref{fig:d4mult}). We implement this on the Rigetti Aspen-9 QPU, whose lattice topology is shown in Fig. \ref{fig:aspen9}. In order to minimize the number of SWAPs necessary to compile the circuit onto the native hardware, we use a 6-qubit sub-lattice consisting of the identifications $\left( a_1, b_1, c_1, a_2, b_2, c_2 \right) = \left( 22, 30, 35, 21, 37, 36 \right)$. This identification ensures only nearest-neighbor interactions in the implementation of the gate. In addition to 2-qubit gates such as CPHASE \cite{Reagor2018} and XY \cite{Abrams2020}, the Rigetti hardware also allows the use of 3-qubit  gates \cite{hill2021realization}. This can be used to compile the Toffoli gate with a single application of the CCPHASE gate, up to a few single-qubit gates.

In order to benchmark the multiplication gate, we start with each possible pair of 3-bitstrings, apply the multiplication gate, and obtain the fraction of correct bitstrings that we measure as output from a total of 10,000 shots. Using only 2-qubit gates to compile the Toffoli in Fig. \ref{fig:d4mult}, we can obtain some depth reduction by identifying a CNOT followed by a SWAP operation with a single XY gate (upto single-qubit gates) as described in \cite{PhysRevA.67.032301}. Using this approach, the average fraction of correct output bitstrings, over all possible input pairs of 3-bitstrings, is found to be $\sim 0.19(6)$ where the standard deviation is reported in parenthesis. However, if we use the native CCPHASE gate to compile the Toffoli in the multiplication gate, the average fraction of correct output bitstrings goes up to $\sim 0.89(18)$. If we instead take the majority vote of 200 successive shots, we boost the average fraction of correct output bitstrings even more to $\sim 0.91(15)$.

\subsection{$D_4$ trace gate and Fourier gate}

\begin{figure*}
	\centering
\includegraphics[width=\textwidth]{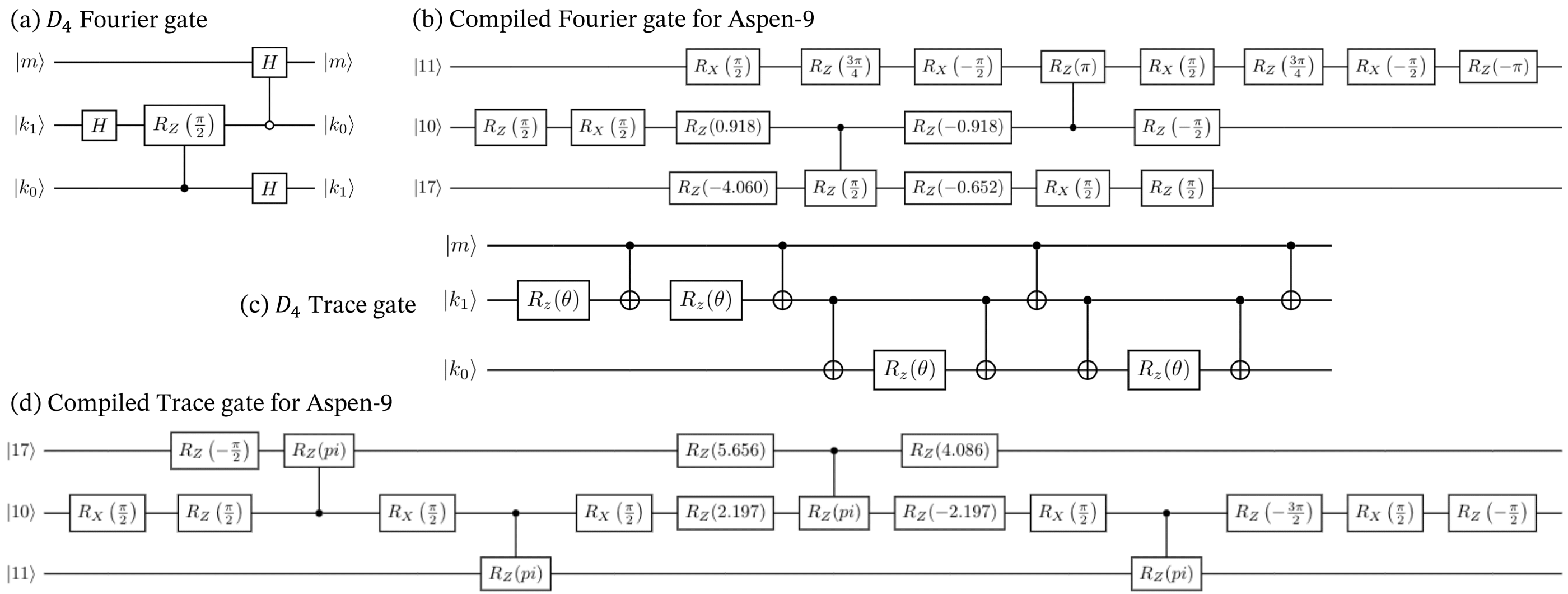}
	\caption{$\mathfrak U_F$ and $\mathfrak U_{\Tr}(\theta=\pi/2)$  for $D_4$ theory and their compiled versions for Aspen-9 QPU with qubits (11, 10, 17). We use the Quilc compiler to convert the circuits to versions using only native gates of Aspen-9 QPU.}
	\label{fig:circuit_trace_Fourier_d4}
\end{figure*}

To benchmark the fidelities of $\mathfrak U_F$ and $\mathfrak U_{\Tr}(\theta=\pi/2)$ of the $D_4$ theory,  we carry out Quantum Process Tomography (QPT) \cite{nielsen_chuang_2010}.
To minimize SWAP gates due to Aspen-9 connectivity, we swap the qubit ordering for a linear connectivity and implement the circuits on the qubits (17, 10, 11) of Aspen-9. We also allow the qubit ordering to be different at the beginning and at the end of the circuits as shown in Fig.~\ref{fig:circuit_trace_Fourier_d4}(a) and (c). The circuits are compiled into the native gate set \{$\mathrm{RZ}(\theta)$, $\mathrm{RX}(k\pi/2)$, CPHASE, CZ, XY\} by Quilc, and need 2 and 4 two-qubit CPHASE gate, respectively as seen in Fig.~\ref{fig:circuit_trace_Fourier_d4}(b) and (d).

\begin{figure*}
	\centering
\includegraphics[width=0.98\textwidth]{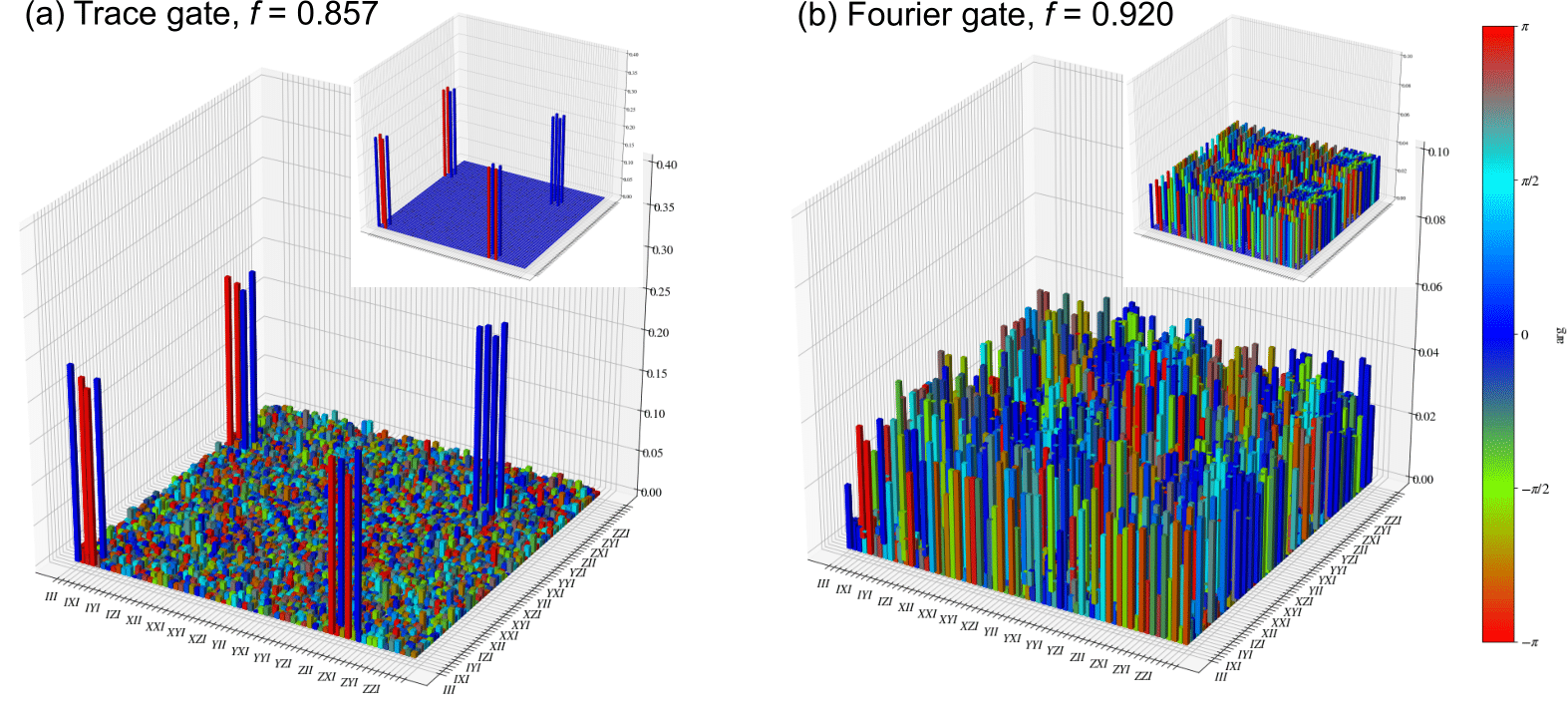}
	\caption{The $\chi$ matrices associated with $\mathfrak U_{\Tr}(\theta=\pi/2)$ and $\mathfrak U_{F}$. The process fidelity is determined by $f = \mathrm{Tr}(\chi_{\mathrm{target}}^{\dagger} \chi)$, where the target $\chi_{\mathrm{target}}$ is computed by noiseless simulator (see insets). $\mathfrak U_{\Tr}(\theta=\pi/2)$ has a lower fidelity $f=0.857$ compared to that of $\mathfrak U_{F}$ being $f=0.920$ since $\mathfrak U_{\Tr}(\theta=\pi/2)$ consists of two more CZ gates. While the process tomography involves pairs of all $4^3$ Pauli operators, to avoid overcrowding, we only display the labels of every four of the operators in the figures above.}
	\label{fig:QPT_Fourier_trace}
\end{figure*}

QPT measures the process fidelities of $\mathfrak U_F$ and $\mathfrak U_{\Tr}(\theta=\pi/2)$ to be $0.920$ and $0.857$. The process infidelity is dominated by the error of the two-qubit CPHASE gates which are calibrated to be around $2\%$ to $3\%$ at the time of the experiments. The $\chi$ matrices measured with 8000 shots are shown in Fig.~\ref{fig:QPT_Fourier_trace} with the inset ideal matrices.
Readout error mitigation is implemented by modeling the readout error as a classical stochastic process characterized by a confusion matrix, which can be determined by preparing all bit strings $|000\rangle$, $|001\rangle$, ..., $|111\rangle$ and measuring the output. Any distribution is then post-processed by inverting the confusion matrix to mitigate the readout error. More details of the readout error mitigation can be found in Refs.\ \cite{peters2021perturbative,Nation2021ScalableReadoutMitigation}.


\subsection{$D_4$ inversion gate}
\begin{figure}[h]
    \includegraphics[width=7cm]{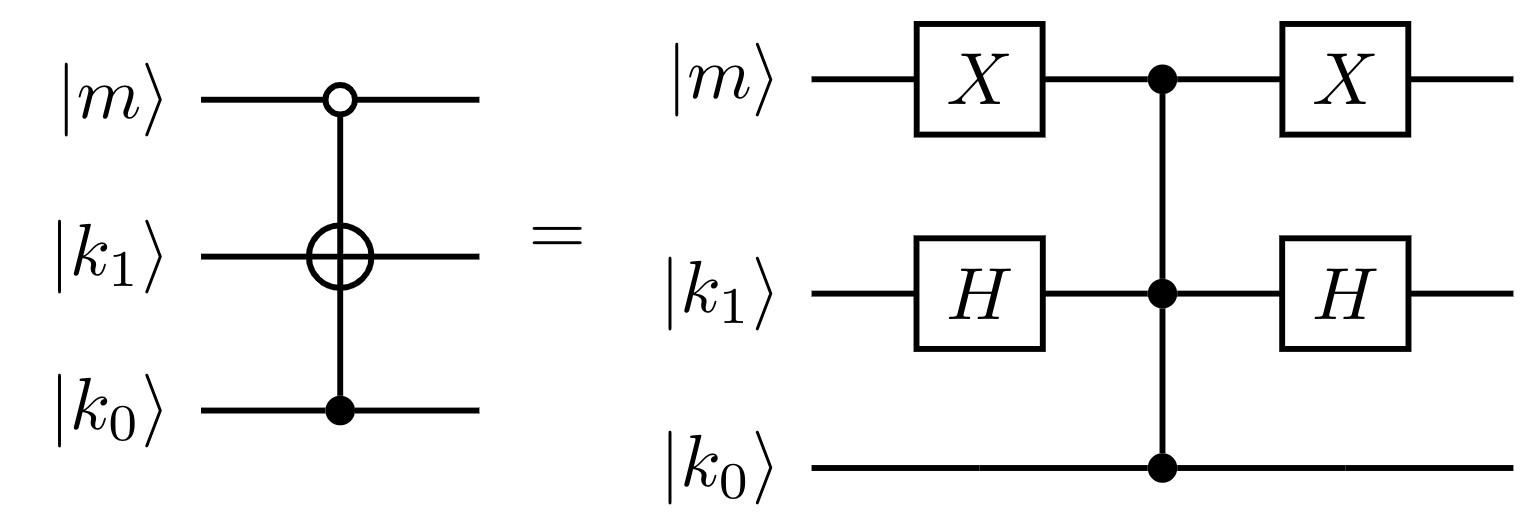}
    \caption{Inversion gate for $D_4$ theory.}
    \label{fig:d4inverse}
\end{figure}

As described in Sec. \ref{sec:inverse}, in order to construct the inversion gate, we need to apply the 2's complement (neglecting the leading bit) of the \textit{rotation} register controlled on the value of the \textit{reflection} qubit being 0. The only non-trivial operations the $D_4$ inversion gate therefore has are $\vert 001\rangle \rightarrow \vert 011 \rangle$ and $\vert 011 \rangle \rightarrow \vert 001 \rangle$. This operation can be implemented using a single CCPHASE($\pi$) gate \cite{hill2021realization}, with a few additional single-qubit gates, as shown in Fig. \ref{fig:d4inverse}. The process fidelity of the CCPHASE gate is computed to be $\sim 87.1\%$ on the Aspen-9 sub-lattice $(10, 11, 12)$ (see Fig. \ref{fig:aspen9}) using cycle benchmarking \cite{cycle-benchmarking}.

\section{\label{sec:Conclusions}Conclusions}
In this paper, we have shown how to construct quantum circuits for the simulation of arbitrary $D_{2^n}$ gauge theories. The operations were shown to reduce to simpler ones such as computing the two's complement, or binary arithmetic, and therefore benefit from the wide variety of techniques used to implement such operations. The Fourier gate was shown to assume a particularly simple form. All these operations were shown to scale as $O(n)$, or as a low-degree polynomial in $n$ in the case of the trace gate,  providing an exponential advantage over classical state vector simulation. Experimentally, we found the success rate of the various operations for $D_4$ theory to be greater than or equal to roughly $80\%$ on Rigetti's Aspen-9 quantum processor. These findings provide encouragement that large scale lattice simulations of gauge theories are within reach.

Looking to the future, several directions warrant mention.  The first would be to extend the construction of primitive gates to gauge theories beyond $D_{2^n}$, in particular to the crystal-like subgroups of $SU(N)$ theories.  The second would be to perform a detailed resource analysis both on the individual gates and algorithms for state preparation~\cite{Harmalkar:2020mpd,Gustafson:2020yfe} and extracting physical observables~\cite{Lamm:2019uyc,Cohen:2021imf} from simulations on specific architecture.  Another followup would investigate the performance of these gates and their combinations on current devices.

\begin{acknowledgments}
We would like to thank the Rigetti team for useful feedback and assistance with running experiments on the Aspen-9 processor, particularly Alex Hill, Mark Hodson, Bram Evert, Nicolas Didier, and Matt Reagor. We are grateful for support from NASA Ames Research Center. 
This material is based upon work supported by the U.S. Department of Energy, Office of Science, National Quantum Information Science Research Centers, Superconducting Quantum Materials and Systems Center (SQMS) under contract number DE-AC02-07CH11359. S.H., and also M.S.A. for work done after the initial posting of this paper on arXiv, were supported by the NASA Academic Mission Services, Contract No. NNA16BD14C. 
Fermilab is operated by Fermi Research Alliance, LLC under contract number DE-AC02-07CH11359 with the United States Department of Energy.
\end{acknowledgments}

\bibliography{refs}

\appendix
\section{Algebraic properties of dihedral groups} 
\label{appendix:algebraic-DN}
Here, we note a few 
important properties of $D_N$, the \textit{dihedral group} of symmetries of a regular $N$-sided polygon, which is generated by two elements: $r$ (a rotation) and $s$ (a reflection) such that $r^N = s^2 = e$, the identity element. Here $N$ can be any positive integer, and 
 it follows that each $D_N$ is isomorphic to the semidirect product of cyclic groups $\mathbb{Z}_N\rtimes\mathbb{Z}_2$. 
Each of the $2N$ elements of $D_N$ can be uniquely expressed as $s^{m}r^{k}$, where $m \in \{0,1\}$ and $k \in \{0,1,\dots, N-1\}$. The two generators 
satisfy the property $srs = r^{-1} = r^{N-1}$, or equivalently $sr = r^{N-1}s$,
which in geometric terms means that a mirror reflection of a rotation 
gives a rotation in the opposite direction. 
Observe that this implies
  %
  %
\begin{eqnarray}
sr^{k+1}= r^{N-1}sr^{k}=\dots=r^{N-(k+1)}s
\end{eqnarray}
so that by induction we have $sr^{k} = r^{N-k}s$ for $k \in \{0, \dots, N-1 \}$. These properties can be summarized as
\begin{equation}
    s^{m}r^{k} = r^{Nm + (-1)^{m}k} s^{m}.
\label{eqn:switch-r-s}
\end{equation}
Through a similar calculation we also have
\begin{equation}
r^{k}s^{m} = s^{m}r^{Nm + (-1)^{m}k}.
\end{equation}
Using the above we find the product rule of Eq. \eqref{eqn-product}:
\begin{eqnarray}
s^{m_1} r^{k_1} \cdot s^{m_2} r^{k_2} &=& s^{m_1} s^{m_2} r^{Nm_2 + (-1)^{m_2} k_1} r^{k_2} \nonumber \\
&=& s^{m_1 + m_2} r^{Nm_2 + (-1)^{m_2}k_1 + k_2}.
\end{eqnarray}

From Eq. \eqref{eqn:switch-r-s} we the inverse of $sr^{k}$ to be
\begin{equation}
    \left(sr^{k} \right)^{-1} = r^{N-k}s = sr^{k}
\end{equation}
while the inverse of $r^{k}$ is simply $\left( r^{k}\right)^{-1} = r^{N-k}$ so that in general, the inverse of a $D_N$ element is given by Eq. \eqref{eqn:inverse}, i.e.,   $\left( s^{m} r^{k} \right)^{-1} =  s^{m} r^{(N-k)(1-m) + mk}$.

\section{Proof that Fourier gate diagonalizes $D_N$ (even $N$) Kinetic gate}
\label{appendix:proof-Fourier-kinetic}
Given $M_{ij} = \operatorname{Re}\left[\mathrm{Tr}\left(\rho^{\dagger}(g_i) \rho(g_j)\right)\right]$ and the matrix $T$ with entries $T_{ij} = \exp{\beta M_{ij}}$, here we show that $FTF^{\dagger}$, where $F$ is the unitary matrix corresponding to the nonabelian Fourier transform, is diagonal. Moreover, we provide an explicit form of this diagonal matrix for arbitrary $D_N$ (even $N$).\\

Note that $M$ is dependent on the representation we use. There are 4 1D irreps of the $D_N$ (even $N$) group,
\begin{itemize}
    \item $\rho_A:$ $r^{k} \rightarrow 1$, $sr^{k} \rightarrow 1$
    \item $\rho_B:$ $r^{k} \rightarrow 1$, $sr^{k} \rightarrow -1$
    \item $\rho_C:$ $s^{m}r^{k} \rightarrow 1$ for even $k$; $s^{m}r^{k} \rightarrow -1$ for odd $k$
    \item $\rho_D:$ $r^{k} \rightarrow 1$, $sr^{k} \rightarrow -1$ for even $k$; and $r^{k} \rightarrow -1$, and $sr^{k} \rightarrow 1$ for odd $k$,
\end{itemize}
where $m\in \{0,1\}$ and $k \in \{0\dots N-1\}$, and $\frac{N}{2}-1$ 2D irreps, 
\begin{eqnarray}
\phi_{(l)}(s^m r^k) = X^m \begin{pmatrix}
e^{i 2\pi l k /N} & 0 \\
0 & e^{-i 2\pi l k /N}
\end{pmatrix}
\end{eqnarray}
with $1 \leq l < N/2$ and $0 \leq k < N$. We work with the $l=1$ 2D irreps, and denote $\phi(s^m r^k) \equiv \phi_{(1)}(s^m r^k)$ in what follows for simplicity. Then, letting $i \equiv (m^{\prime}, k^{\prime}) = Nm^{\prime} + k^{\prime}$, $j \equiv (m,k) = Nm+k$, it is clear that $M_{ij}$ can be non-zero only when $m=m^{\prime}$. In this case, we see that
\begin{eqnarray}
\phi^{\dagger}(s^m r^{k^{\prime}}) \phi(s^m r^k) = \begin{pmatrix}
e^{i 2 \pi (k - k^{\prime})/N} & 0 \\
0 & e^{-i 2 \pi (k - k^{\prime})/N}
\end{pmatrix} \nonumber \\
\end{eqnarray}
so that $M_{i;j} = M_{(m^{\prime},k^{\prime});(m,k)} = 2 \delta_{m, m^{\prime}} \cos{ \left[ 2\pi \left( \frac{k^{\prime} - k}{N} \right)\right]}$. Therefore,
\begin{eqnarray}
T_{ij} \equiv T_{(m',k');(m,k)} &=& e^{2\delta_{m,m'}\beta \cos{\left[2\pi (k' - k)/N \right]}} \nonumber \\
&&
\end{eqnarray}


Now the Fourier matrix is built out of the inequivalent irreps of the $D_N$ group, and can be represented as
\begin{eqnarray}
F &=& \begin{pmatrix}
\rho_a \\
\rho_b \\
\rho_c \\
\rho_d \\
\phi_{00(1)} \\
\phi_{01(1)} \\
\phi_{10(1)} \\
\phi_{11(1)} \\
\vdots \\
\phi_{00(N/2 - 1)} \\
\phi_{01(N/2 - 1)} \\
\phi_{10(N/2 - 1)} \\
\phi_{11(N/2 - 1)}
\end{pmatrix}
\end{eqnarray}
so that we have
\begin{eqnarray}
F_{0i} &= \rho_a &= [1]^{2N} \nonumber \\
F_{1i} &= \rho_b &= [1]^{N} [-1]^{N} \nonumber \\
F_{2i} &= \rho_c &= [1,-1]^{N} \nonumber \\
F_{3i} &= \rho_d &= [1,-1]^{N/2} [-1,1]^{N/2}
\end{eqnarray}
where $[a]^{m_1} [b]^{m_2}$ denotes an $m_{1} + m_{2}$-dimensional row vector with the first $m_1$ entries equaling $a$, and the next $m_2$ entries equaling $b$, and $[a,b]^{m}$ denotes a $2m$-dimensional row vector with entries alternating between $a$ and $b$. Let $F^{\prime} = FT$ and $\rho_{a}^{\prime}, \dots, \rho_{d}^{\prime}, \phi_{00(1)}^{\prime}, \dots, \phi_{11(N/2-1)}^{\prime}$ denote its rows.

We then have
\begin{eqnarray}
(FT)_{0;(m,k)} &=& \sum_{m', k'} F_{0;(m',k')} T_{(m',k');(m,k)} \nonumber \\
&=& N + \sum_{k'=0}^{N-1} e^{2\beta \cos{\left[ 2\pi (k' - k)/N\right]}} \nonumber \\
&=& N + \sum_{k'=0}^{N-1} e^{2\beta \cos{\left( 2\pi k'/N\right)}}
\label{appdx-eqn:rho-a-prime}
\end{eqnarray}
where in the last step we have repeatedly used the identity $\cos{\left( \frac{2\pi(N-k)}{N}\right)} = \cos{\left( \frac{2\pi k}{N}\right)}$. The last expression in Eq. \ref{appdx-eqn:rho-a-prime} is independent of $(m,k)$, so that
\begin{eqnarray}
\rho_{a}^{\prime} = \left(N + \sum_{k'=0}^{N-1} e^{2\beta \cos{\left( 2\pi k'/N\right)}} \right) \rho_a
\end{eqnarray}
Similarly,
\begin{eqnarray}
(FT)_{1;(m,k)} &=& F_{1;(m',k')} T_{(m',k');(m,k)} \nonumber \\
&=& \sum_{k'} e^{2\delta_{0,m}\beta \cos{\left[2\pi (k' - k)/N \right]}} \nonumber \\
&& \quad - \sum_{k'} e^{2\delta_{1,m} \beta \cos{\left[ 2\pi (k' - k)/N\right]}} \nonumber \\
&=& (1 - 2m) \left( \sum_{k'} e^{2\beta \cos{(2\pi k'/N)}} - N \right) \nonumber \\
&& 
\end{eqnarray}
and since $(\rho_{b})_{(m,k)} = (1-2m)$, we have
\begin{eqnarray}
\rho_{b}^{\prime} = \left( \sum_{k'} e^{2\beta \cos{(2\pi k'/N)}} - N \right) \rho_b
\end{eqnarray}
Through very similar calculations, we find
\begin{eqnarray}
(FT)_{2;(m,k)} &=& \sum_{m'} \left[ \sum_{k' \text{ odd}} (-1) e^{2\delta_{m,m'} \beta \cos{\left[ 2\pi (k' -k)/N \right]}} + \right. \nonumber \\
&& \; \left. \sum_{k' \text{ even}} e^{2\delta_{m,m'} \beta \cos{\left[ 2\pi (k' -k)/N \right]}} \right] \nonumber \\
&=& \sum_{k'} (-1)^{k'} e^{2\beta \cos{\left( 2\pi k'/N \right)}}
\end{eqnarray}
and, for the last of the 1D irreps,
\begin{eqnarray}
(FT)_{3;(m,k)} &=& (-1)^{m} \sum_{k'} (-1)^{k'} e^{2\beta \cos{\left( 2\pi k'/N\right)}}
\end{eqnarray}
so that
\begin{eqnarray}
\rho_{c,d}^{\prime} &=& \left( \sum_{k'} (-1)^{k'} e^{2\beta \cos{\left( 2\pi k'/N\right)}} \right) \rho_{c,d}
\end{eqnarray}
Next, for the 2D irreps, we have
\begin{eqnarray}
\phi_{00(l)}(s^m r^k) &=& (1-m) e^{i 2\pi l k /N} \nonumber \\
\phi_{01(l)}(s^m r^k) &=& m e^{-i 2\pi l k /N} \nonumber \\
\phi_{10(l)}(s^m r^k) &=& m e^{i 2\pi l k/N} \nonumber \\
\phi_{11(l)}(s^m r^k) &=& (1-m) e^{-i 2\pi l k/N}
\end{eqnarray}
We will make use of the identities
\begin{eqnarray}
\sum_{k'=0}^{N-1} \cos{\left( \frac{2\pi l k'}{N}\right)} &=& \sin{(\pi l)} \left(\cos{(\pi l) \cot{\left(\frac{\pi l}{N} \right)} + \sin{(\pi l)}} \right) \nonumber \\
&& \\
\sum_{k'=0}^{N-1} \sin{\left( \frac{2\pi l k'}{N}\right)} &=& \frac{1}{2} \left( \cos{\left( \frac{\pi l}{N}\right)} - \cos{\left( \frac{\pi l (2N-1)}{N}\right)}\right) \nonumber \\
&& \qquad\qquad \times\;\; \text{cosec}\left(\frac{\pi l}{N} \right)
\label{appdx-eqn:cos-sin-sum}
\end{eqnarray}
both of which vanish for $l \in \mathbb{Z}$. Therefore,
\begin{eqnarray}
\sum_{k'=0}^{N-1} e^{\pm i 2 \pi l k' / N} &=& 0
\label{appdx-eqn:exp-sum-vanishes}
\end{eqnarray}
for $1 \leq l < N/2$. We will also make use of the identity
\begin{eqnarray}
\sum_{k'=0}^{N-1} e^{\pm i 2 \pi l k'/N} &=& \sum_{k'=0}^{N-1} e^{\pm i 2 \pi l (k' + k) /N}
\label{appdx-eqn:exp-period-shift}
\end{eqnarray}
for $k = 0, \dots, N-1$.\\

Now, analyzing the transformation of the 2D irreps as we did before for the 1D irreps, we have
\begin{eqnarray}
\phi_{00(l)}^{\prime}(s^m r^k) &=& \sum_{(m',k')} \phi_{00(l)}(s^{m'}r^{k'}) T_{(m',k');(m,k)} \nonumber \\
&=& \sum_{m',k'} (1-m') e^{i2 \pi l k'/N} e^{2\beta \delta_{m,m'} \cos{[2\pi (k' - k)/N]}} \nonumber \\
&=& \begin{cases}
\sum_{k'} e^{i 2 \pi l k'/N} e^{2\beta \cos{(2\pi k'/N)}} & \text{, $m=0$}\\
\sum_{k'} e^{i 2 \pi l k'/N} & \text{, $m=1$}
\end{cases} \nonumber \\
&=& (1-m) e^{i 2 \pi l k/N} \sum_{k'} e^{i 2\pi l k'/N} e^{2\beta \cos{(2 \pi k'/N)}} \nonumber \\
&=& \left( \sum_{k'} e^{i 2\pi l k'/N} e^{2\beta \cos{(2 \pi k'/N)}} \right) \phi_{00(l)}
\end{eqnarray}
where in the second to last equality, we have used Eqns. \ref{appdx-eqn:exp-sum-vanishes} and \ref{appdx-eqn:exp-period-shift}. Repeating essentially the same arguments, we also obtain
\begin{eqnarray}
\phi_{01(l)}^{\prime}(s^m r^k) &=& \left( \sum_{k'} e^{-i 2\pi l k'/N} e^{2\beta \cos{(2 \pi k'/N)}} \right) \phi_{01(l)} \nonumber \\
\phi_{10(l)}^{\prime}(s^m r^k) &=& \left( \sum_{k'} e^{i 2\pi l k'/N} e^{2\beta \cos{(2 \pi k'/N)}} \right) \phi_{01(l)} \nonumber \\
\phi_{11(l)}^{\prime}(s^m r^k) &=& \left( \sum_{k'} e^{-i 2\pi l k'/N} e^{2\beta \cos{(2 \pi k'/N)}} \right) \phi_{11(l)} \nonumber \\
&&
\end{eqnarray}

Altogether, we have shown that $\rho_{i}^{\prime} \propto \rho_{i}$ for all irreducible representations $\rho_i$. Finally, using the Schur orthogonality relations which state that for two inequivalent irreps of some finite group G, $\phi: G \rightarrow U_{n}(\mathbb{C})$ and $\rho: G \rightarrow U_{m}(\mathbb{C})$, we have
\begin{equation}
    \langle \phi_{kl}, \rho_{ij} \rangle = 0,\hspace{0.4cm}\langle \phi_{kl}, \phi_{ij}\rangle = \delta_{ik} \delta_{jl}
\end{equation}
and so we find
\begin{widetext}
\begin{eqnarray}
FTF^{\dagger} &=& \text{Diag} \left\lbrace \left(N + \sum_{k'=0}^{N-1} e^{2\beta \cos{\left( 2\pi k'/N\right)}} \right), \left( \sum_{k'} e^{2\beta \cos{(2\pi k'/N)}} - N \right), \left[ \sum_{k'} (-1)^{k'} e^{2\beta \cos{\left( 2\pi k'/N\right)}}  \right]^2, \right. \nonumber \\
&& \left. \qquad \left[ \left( \sum_{k'} e^{i 2\pi l k'/N} e^{2\beta \cos{(2 \pi k'/N)}} \right), \left( \sum_{k'} e^{-i 2\pi l k'/N} e^{2\beta \cos{(2 \pi k'/N)}} \right), \right. \right. \nonumber \\
&& \qquad \qquad \left. \left. \left( \sum_{k'} e^{i 2\pi l k'/N} e^{2\beta \cos{(2 \pi k'/N)}} \right), \left( \sum_{k'} e^{-i 2\pi l k'/N} e^{2\beta \cos{(2 \pi k'/N)}} \right) \right]_{l=1}^{\frac{N}{2}-1} \right\rbrace
\end{eqnarray}
\end{widetext}
where, $[f_{1}(l), \dots, f_{k}(l)]_{l=1}^{m}$ denotes $mk$ entries with increasing values of $l$. The expression above is valid for even $N$. We do not consider the odd $N$ case in our paper, though the analysis may be carried out similarly.

\end{document}